\documentclass[11pt]{article}

\usepackage{epsfig,subfigure}
\usepackage{amssymb, amsmath}
\usepackage{color}
\newcommand{\define}{\stackrel{\triangle}{=}}
\def\QED{\mbox{\rule[0pt]{1.5ex}{1.5ex}}}
\def\proof{\noindent\hspace{2em}{\it Proof: }}

\usepackage{graphicx}
\usepackage{color}

\usepackage{amssymb}
\usepackage{amsmath,amsfonts,amssymb}
\usepackage{verbatim}
\usepackage{stfloats}
\usepackage[bookmarks=false]{}

\newtheorem{theorem}{Theorem}
\newtheorem{corollary}{Corollary}

\newtheorem{lemma}{Lemma}

\newtheorem{example}{Example}

\begin{document}
\date{}
\title{On the Optimality of Treating Interference as Noise\\ for $K$ user Parallel Gaussian Interference Networks
}
\author{ \normalsize Hua Sun and Syed A. Jafar \\
{\small Center for Pervasive Communications and Computing (CPCC)}\\
{\small University of California Irvine, Irvine, CA 92697}\\
{\small \it Email: \{huas2, syed\}@uci.edu}
}
\maketitle

\begin{abstract}
It has been shown recently by Geng et al. that in a $K$ user Gaussian interference network, if  for each user the desired signal strength is no less than the sum of the strengths of the strongest interference from this user and the strongest interference to this user (all signal strengths measured in dB scale), then power control and treating interference as noise (TIN) is sufficient to achieve the entire generalized degrees of freedom (GDoF) \emph{region}. Motivated by the intuition that the deterministic model of Avestimehr et al. (ADT deterministic model) is particularly suited for exploring the optimality of TIN, the results of Geng et al. are first re-visited under the ADT deterministic model, and are shown to directly translate between the Gaussian and deterministic settings. Next, we focus on the extension of these results to parallel interference networks, from a sum-capacity/sum-GDoF perspective. To this end, we interpret the explicit characterization of the sum-capacity/sum-GDoF of a TIN optimal network (without parallel channels) as a minimum weighted matching problem in combinatorial optimization, and obtain a  simple characterization in terms of a partition of the interference network into vertex-disjoint cycles. Aided by insights from the cyclic partition, the  sum-capacity optimality of TIN for $K$ user parallel  interference networks is characterized for the ADT deterministic model, leading ultimately to corresponding GDoF results for the Gaussian setting. In both cases, subject to a \emph{mild} invertibility condition the optimality of TIN is shown to extend to parallel networks in a separable fashion.

\end{abstract}
\newpage

\section{Introduction}

Treating interference as noise (TIN) is a strategy that is universally applied in wireless networks to deal with interference from users that are far away. Interestingly, it is also known to be capacity optimal when the interference is sufficiently weak \cite{Etkin_Tse_Wang, Shang_Kramer_Chen, Motahari_Khandani, Sreekanth_Veeravalli, Geng_TIN}. Most relevant to this work is the recent result by Geng et al. in \cite{Geng_TIN}, where a  broadly applicable condition is identified and shown to be sufficient (also conjectured to be necessary in almost all cases) for TIN to achieve the generalized degrees of freedom (GDoF) region. The GDoF optimality of TIN then serves as a stepping stone to a further tightening of the result, so that whenever Geng et al.'s condition holds, TIN is shown to achieve the entire capacity region within a constant gap. 

Geng et al.'s result highlights the advantage of the GDoF metric for obtaining finer insights into the capacity of wireless networks, relative to the more widely studied degrees of freedom (DoF) metric. While DoF studies have contributed a number of fundamental insights, the DoF metric is limited in that it treats all non-zero channels as essentially equally strong (capable of carrying exactly 1 DoF). Thus, insights into schemes such as TIN, which rely very much on certain signals being much weaker than others, cannot be obtained directly from DoF studies. The GDoF perspective is crucial for such insights, and serves as the logical next step after DoF in the pursuit of capacity through progressively refined approximations. The advantage of the GDoF metric is amply evident in the study of the 2 user interference network by Etkin et al. in \cite{Etkin_Tse_Wang}, where the DoF metric only provides a trivial answer, whereas the GDoF metric identifies all of the important operational regimes, leading ultimately to a  characterization of the entire capacity region within a 1 bit gap. 

The richness of the GDoF metric naturally comes at the cost of reduced tractability, especially since even the simpler DoF metric is far from fully understood for wireless networks. As such GDoF characterizations are few and far in between \cite{Etkin_Tse_Wang, Parker_Bliss_Tarokh, Jafar_Vishwanath_GDOF, Gou_Jafar_O1, Huang_Cadambe_Jafar, Karmakar_Varanasi}. This motivates simpler alternatives such as the ADT deterministic model of \cite{Avestimehr_Diggavi_Tse, Bresler_Tse, Bresler_Parekh_Tse}. The ADT deterministic model captures much of the essence of the GDoF framework  --- the diversity of signal strengths --- but is less useful when the finer details such as the  channel phase or the distinction between rational and irrational realizations become critical. Unfortunately, since these finer details are important for wireless interference networks with 3 or more users (even from a DoF perspective) \cite{Etkin_Ordentlich, Motahari_Gharan_Khandani, Zamanighomi_Wang, Cadambe_Jafar_Wang}, the ADT deterministic model has found limited use in such settings. 

The main idea motivating this work is that while the ADT deterministic model may not be suitable for studying the more fragile regimes, it could still be well suited for studying those robust regimes where the finer aspects of channel realizations are not relevant. Given this insight, and since the regime where TIN is optimal is arguably the most robust regime, it follows that the ADT deterministic model should suffice to identify this regime in the GDoF sense and to study its properties. As initial verification of this insight, we begin by exploring the TIN optimality result of Geng et al. in the ADT framework. Indeed, the optimality conditions and the GDoF region are not only easily mapped to the ADT deterministic model, but also become  more transparent in the deterministic setting. Encouraged by this insight, we proceed to the main contribution of this work --- exploring the  optimality of TIN for $K$ user parallel Gaussian interference networks. 

Optimality of TIN for parallel Gaussian interference networks is an intriguing question for the following reasons. On the one hand, with  the exception of the MAC-Z-BC network (which contains the multiple access channel, Z-channel and broadcast channel as special cases), it is known that all parallel Gaussian networks are in general inseparable \cite{Cadambe_Jafar_inseparable, Cadambe_Jafar_MACZBC, Lalitha_inseparable}. The benefits of joint coding across parallel channels can be quite substantial and extend all the way from higher DoF \cite{Cadambe_Jafar_inseparable} to simple achievable schemes and near-optimal rates at finite SNR \cite{Nazer_Gastpar_Jafar_Vishwanath, Jafar_Ergodic}. On the other hand, for the 2 user interference network, extensions to parallel channels have been made from an exact \emph{sum-capacity} perspective in \cite{Shang_ParallelNoisy} and from a GDoF perspective in \cite{Karmakar_Varanasi}\footnote{Parallel interference networks may be seen as a special case of MIMO interference networks.}. In both cases, the results support separability of TIN optimal sub-channels. However, the insights from the $2$ user setting do not directly extend to the $K$ user interference network. For example,  the GDoF region for the TIN optimal 2 user interference network is easily seen to be  polymatroidal, whereas the GDoF region of TIN optimal $K$ user interference networks, with $K\geq 3$, is no longer polymatroidal.
The distinction is particularly significant for parallel channels. The GDoF region of 2 user TIN optimal parallel interference networks is  simply the  direct sum of the corresponding sum-rate bounds for all  the sub-channels \emph{and} is  achieved by separate TIN on each sub-channel. This is in general not the case with $3$ or more users (a simple example is provided in Section \ref{sec:SFregion}).  Given the significant challenges in going beyond $2$ users, it is most intriguing if the separability of parallel Gaussian interference networks will hold in the regime where TIN is sum-GDoF optimal. In other words, if each of the sub-channels of a $K$ user interference network satisfies the TIN optimality condition of Geng et al., then will TIN continue to be sum-GDoF optimal for the parallel channel setting?

The focus on sum-GDoF motivates us to first seek a more explicit characterization. To this end, we show that the sum-GDoF characterization for a $K$ user interference network is essentially a minimum weighted matching problem in combinatorial optimization. Consequently, the sum-GDoF are characterized in terms of a partition of the interference network into disjoint cycles. Aided by the insights from the cyclic partition approach, we explore the sum-capacity optimality of TIN for $K$ user parallel deterministic interference networks under the ADT deterministic model. A separable extension of the optimality of TIN to parallel interference networks is obtained subject to a mild invertibility condition. The result is then translated into the GDoF framework for parallel Gaussian interference networks. In terms of answering the main question, the implication is that if each of the sub-channels satisfies the TIN optimality condition of Geng et al., then subject to a mild invertibility condition, a separate TIN scheme for each sub-channel continues to be sum-GDoF optimal for the overall $K$ user parallel Gaussian interference networks.

\section{System Model, Definitions, and Notation}\label{sec:systemmodel}
\subsection{Gaussian Interference Network Model}
Consider the $K$ user real Gaussian interference network, with $M$ parallel sub-channels, described as
\begin{equation}
\label{original}
{\bf Y}_k(t) = \sum_{i=1}^{K}\tilde{\bf H}_{ki}\tilde{\bf X}_i(t) + {\bf Z}_k(t),~~~\forall k \in [K] \triangleq \{1,2,\ldots,K\},
\end{equation}
where over the $t$-th channel use,
\begin{eqnarray}
{\bf Y}_k(t)  &=& \left[ Y_k^{[1]}(t), Y_k^{[2]}(t), \ldots, Y_k^{[M]}(t) \right]^T \\
\tilde{\bf X}_i(t)  &=& \left[ \tilde{X}_i^{[1]}(t), \tilde{X}_i^{[2]}(t), \ldots, \tilde{X}_i^{[M]}(t) \right]^T
\end{eqnarray}
are the vectors containing the received signals observed at Receiver $k$ and the transmitted symbols from Transmitter $i$, respectively,  and
\begin{eqnarray}
{\tilde{\bf H}_{ki} =  \left[ \begin{array}{cccc}  \tilde h_{ki}^{[1]} & 0 & \ldots & 0\\
    0 & \tilde h_{ki}^{[2]} & \ldots & 0\\
    \vdots & \vdots & \ddots & \vdots\\
     0 & 0 & \cdots  & \tilde h_{ki}^{[M]}\\
    \end{array}\right]} \label{h}
\end{eqnarray}
is a diagonal channel matrix comprised of the channel coefficients from Transmitter $i$ to Receiver $k$. The superscript within the square parentheses represents  the sub-channel index, $m \in [M] \triangleq \{1,2,\ldots,M\}$. All  channel coefficients are fixed across channel uses. Perfect channel knowledge is available at all transmitters and receivers. The AWGN vector at Receiver $k$ over the $t$-th channel use,
\begin{eqnarray}
{\bf Z}_k(t)  &=& \left[ Z_k^{[1]}(t), Z_k^{[2]}(t), \ldots, Z_k^{[M]}(t) \right]^T
\end{eqnarray}
has zero mean and covariance matrix ${\bf I}_{M}$, where ${\bf I}_M$ represents the $M \times M$ identity matrix. Noise processes are i.i.d over time. All symbols are real.

At Transmitter $i$, an independent message $W_i$ uniformly distributed over the message index set $\{1,2,\ldots,\lceil 2^{nR_i}\rceil\}$ is mapped to the transmitted codeword $[ \tilde{\bf X}_i(1), \tilde{\bf X}_i(2), \ldots, \tilde{\bf X}_i(n) ]$ (abbreviated as $\tilde{\bf X}_i^n$) over $n$ channel uses, and is subject to the average power constraint,
\begin{eqnarray}
\frac{1}{n}\sum_{t=1}^{n}\sum_{m=1}^{M} \mathbb{E}\left| \tilde{X}_i^{[m]}(t) \right|^2 \leq P_i
\end{eqnarray}
where the expectation is over the messages.

At Receiver $k$, the received signal $ [ {\bf Y}_k(1), {\bf Y}_k(2), \ldots, {\bf Y}_k(n) ]$ (abbreviated as $ {\bf Y}_k^n$) is used to produce the estimate $\hat{W}_k$ of the message $W_k$. The probability of error for Receiver $k$ is given by the probability that $\hat{W}_k$ is not equal to $W_k$. A rate tuple $(R_1, R_2, \ldots, R_K)$ is said to be achievable if we have an encoding and decoding mapping such that the probability of error for each  receiver approaches zero as $n$ approaches infinity. The capacity region $\mathcal{C}$ is the closure of the set of all achievable rate tuples. The sum-capacity is defined as $\mathcal{C}_{\Sigma} = \max_{\mathcal{C} } \sum_{k=1}^K R_k$. 

\subsection{GDoF Framework}
Following  \cite{Geng_TIN}, we  now translate the channel model (\ref{original}) into an equivalent normalized form to facilitate GDoF studies. For such a purpose, we define $\tilde{X}_i^{[m]}(t) = \sqrt{P_i}{X}_i^{[m]}(t)$. Then over the $t$-th channel use, the received signal for Receiver $k$ across the $m$-th sub-channel is described by
\begin{equation}
\label{now}
Y_k^{[m]}(t) = \sum_{i=1}^{K} \tilde h_{ki}^{[m]}  \sqrt{P_i} {X}_i^{[m]}(t) + {Z}_k^{[m]}(t).
\end{equation}

Further, we take $P >1$ as a nominal power value, and define
\begin{equation}
\alpha_{ki}^{[m]} \triangleq \left( \frac{\log \left( \left| \tilde h_{ki}^{[m]} \right|^2 P_i \right)} {\log P} \right)^+.\footnote{As noted in \cite{Geng_TIN},  avoiding negative $\alpha$'s, will not influence the GDoF results.}
\end{equation}

The channel model (\ref{now}) becomes
\begin{eqnarray}\label{model}
Y_k^{[m]}(t) &=& \sum_{i=1}^{K} \mbox{sign}(\tilde h_{ki}^{[m]}) \sqrt{P^{\alpha_{ki}^{[m]}}}  {X}_i^{[m]}(t) + {Z}_k^{[m]}(t) \\
&=& \sum_{i=1}^{K} h_{ki}^{[m]}  {X}_i^{[m]}(t) + {Z}_k^{[m]}(t) \label{use}
\end{eqnarray}
where $h_{ki}^{[m]} \triangleq  \mbox{sign}(\tilde h_{ki}^{[m]}) \sqrt{P^{\alpha_{ki}^{[m]}}} $ is the effective channel coefficient and $ {X}_i^{[m]}(t)$ is the equivalent channel input whose power is absorbed into the channel,
\begin{eqnarray}
\frac{1}{n}\sum_{t=1}^{n}\sum_{m=1}^{M} \mathbb{E} \left| {X}_i^{[m]}(t) \right|^2 \leq 1.
\end{eqnarray}
As in \cite{Geng_TIN}, we call $\alpha_{ki}^{[m]}$ the channel strength level. The equivalent model (\ref{use}) will be used in the rest of this paper.

We define the GDoF region as
\begin{eqnarray}
\mathcal{D} \triangleq \left\{ (d_1, d_2,\ldots, d_K) : d_i = \lim_{P \rightarrow \infty} \frac{R_i}{\frac{1}{2} \log P}, \forall i \in \{1,2,\ldots,K\}, (R_1, R_2, \ldots, R_K) \in \mathcal{C} \right\}.
\end{eqnarray}

The sum-GDoF value is defined as $\mathcal{D}_{\Sigma} = \max_{\mathcal{D} } \sum_{k=1}^K d_k$.

\subsection{ADT Deterministic Interference Network Model}
As in the Gaussian case, there are $K$ transmitter-receiver pairs in the ADT deterministic interference network model. Each transmitter wants to communicate with its corresponding receiver. The signal sent from Transmitter $i$, as observed at Receiver $k$, over the $m$-th sub-channel, is scaled up by a nonnegative integer value $n_{ki}^{[m]} \triangleq  \lfloor  \log_2 |h_{ki}^{[m]}| \rfloor = \lfloor  \frac{1}{2}\alpha_{ki}^{[m]}  \log_2 P \rfloor$. 

The channel may be written as
\begin{eqnarray}\label{det}
Y_k^{[m]} &=& \lfloor 2^{n_{k1}^{[m]}} {X}_1^{[m]} \rfloor \oplus \lfloor 2^{n_{k2}^{[m]}} {X}_2^{[m]} \rfloor \oplus \cdots \oplus \lfloor 2^{n_{kK}^{[m]}} {X}_K^{[m]} \rfloor
\end{eqnarray}
where addition is performed on each bit (modulo two). The time index is omitted for compactness. We assume the real-valued channel input is positive and has peak power constraint 1, then it can be written in base 2 as 
\begin{eqnarray}
X_i^{[m]} = 0.X_{i,(1)}^{[m]}X_{i,(2)}^{[m]}X_{i,(3)}^{[m]} \ldots. 
\end{eqnarray}
The capacity region and the associated notions are defined similar to those in the Gaussian setting.

The following directed graph representation will be useful to efficiently present the results in this work.
\subsection{Weighted Directed Graph Representation} 
\begin{figure}[h]
\center
\includegraphics[width=2.5in]{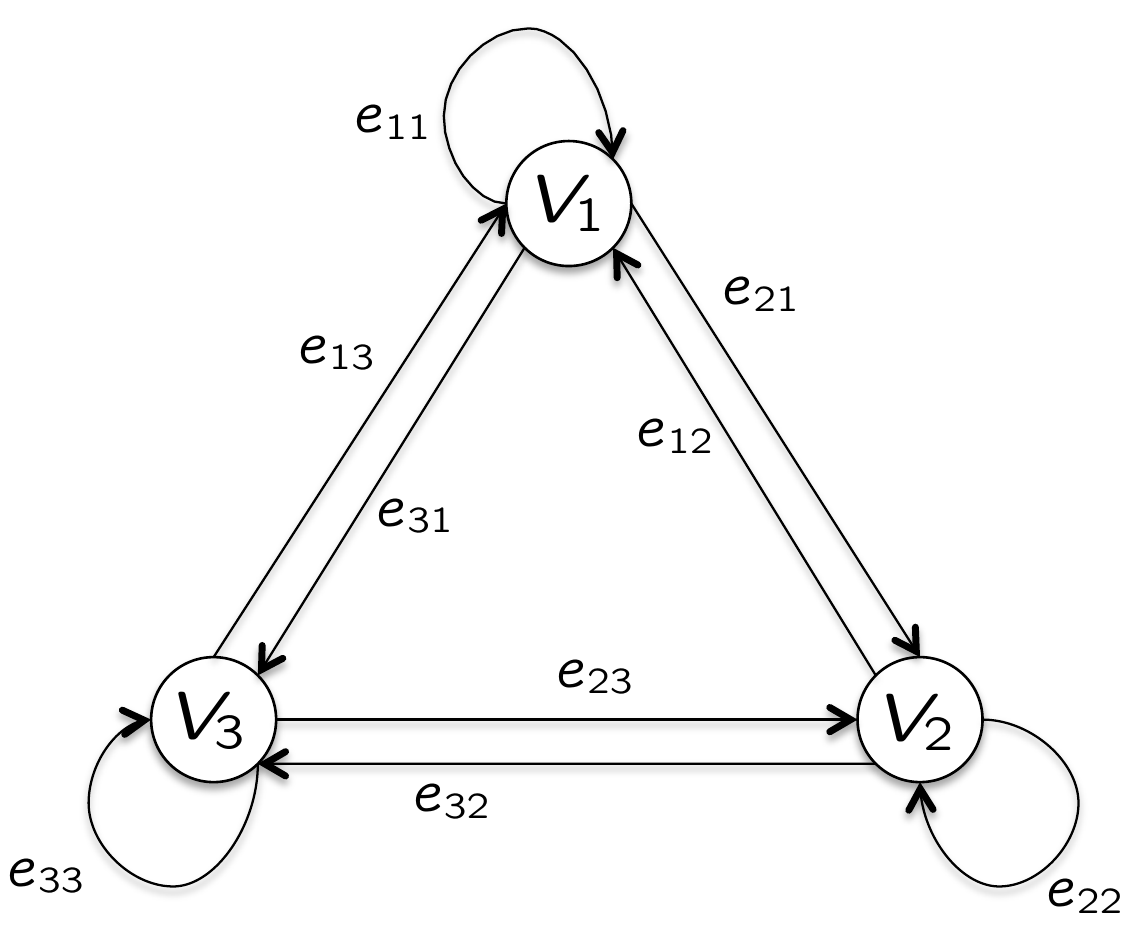}
\caption{\small The directed graph representation of a 3 user interference network.}
\label{dir}
\end{figure}

The directed graph representation of the $K$ user interference network consists of $K$ vertices, $V_1, V_2, \cdots, V_K$, one for each user. Since the vertices correspond directly to users, we will also refer to them as users. For all $(i,j)\in[K]\times[K]$, there is a directed edge $e_{ij}$ from user $j$ to user $i$, with weight $w(e_{ij})$ defined as follows:
\begin{eqnarray}
w(e_{ij})&=&\left\{
\begin{array}{cc}
\alpha_{ij}& \mbox{ if } i\neq j,\\
0 &\mbox{ if } i=j
\end{array}
\right.
\end{eqnarray}
The directed graph for $K=3$ is illustrated in Fig. \ref{dir}.
The directed graph is similarly defined for the ADT deterministic model, with all $\alpha_{ij}$ values replaced by $n_{ij}$ values. 

We are particularly interested in the notion of cycles on this directed graph. We define a cycle, $\pi$,  as a cyclically ordered subset of users, without repetitions.  The set of all cycles is denoted as $[\Pi]$. The cardinality of a cycle, denoted as $|\pi|$ is the number of users that it involves. 
\begin{eqnarray}
|\pi|&=&\sum_{V_k\in\pi} 1, ~~~\forall \pi \in [\Pi]
\end{eqnarray}
A cycle with only one user is a trivial cycle. Two cycles $\pi_p, \pi_q$, are said to be disjoint if they contain no common user, denoted as $\pi_p\cap\pi_q=\phi$.

Introducing a slight abuse of notation in the interest of conciseness, the same cycle, $\pi$,  can also be equivalently represented as a  set of edges representing a closed path where no user is visited more than once. The  weight of a cycle, denoted as $w(\pi)$,  is the sum of the weights of all the edges traversed in completing the cycle.

\begin{eqnarray}
w(\pi)&=&\sum_{e_{ij}\in\pi}w(e_{ij}), ~~~\forall \pi\in[\Pi]
\end{eqnarray}
Note that the weight of a trivial cycle is zero. Intuitively, the weight of a cycle is the accumulation of the strengths of interference terms encountered in the cycle.

As an example, consider the 3 user interference network, for which we have a total of 8 possible cycles, so that
\begin{eqnarray}
[\Pi]&=&\{\{1\}, \{2\}, \{3\}, \{1,2\}, \{2,1\}, \{1,3\}, \{3,1\}, \{2,3\}, \{3,2\}, \{1,2,3\}, \{3,2,1\}\}\\
&=&\{\{e_{11}\}, \{e_{22}\}, \{e_{33}\},\{e_{12},e_{21}\},\{e_{13},e_{31}\},\{e_{23},e_{32}\},\{e_{12},e_{23},e_{31}\},\{e_{32},e_{21},e_{13}\}\}
\end{eqnarray}
\begin{eqnarray}
w(\{1,2,3\})&=&\alpha_{12}+\alpha_{23}+\alpha_{31}\\
&=&w(\{e_{12},e_{23},e_{31}\})
\end{eqnarray}

{\bf Cyclic Partition:} A subset of the set of all cycles, $\Pi\subset [\Pi]$, is said to be a cyclic partition if 
\begin{eqnarray}
\pi_p\cap\pi_q&=&\phi, ~~\forall \pi_p,\pi_q\in\Pi\\
\sum_{\pi\in\Pi}|\pi|&=&K
\end{eqnarray}
In other words, a cyclic partition is a disjoint cyclic cover of the $K$ users.

{\bf Cyclic Partition Bound:} For any cyclic partition $\Pi$, define the corresponding cyclic partition bound, $\mathcal{D}^\Pi_\Sigma$,  as
\begin{eqnarray}
\sum_{k=1}^K d_k&\leq& \sum_{k=1}^K\alpha_{kk}-w(\Pi)
\end{eqnarray}
where 
\begin{eqnarray}
w(\Pi) &=&\sum_{\pi\in\Pi}w(\pi)
\end{eqnarray}
is the net weight of the cyclic partition, representing the total interference encountered in this partition.

Since there are many cyclic partitions, each of which gives rise to a cyclic partition bound, let us denote the tightest of these bounds as the \emph{best cyclic partition bound}, $\mathcal{D}^{\Pi*}_\Sigma$. In the deterministic setting, a cyclic partition bound is denoted by $\mathcal{C}^\Pi_\Sigma$ and the best cyclic partition bound is denoted by $\mathcal{C}^{\Pi*}_\Sigma$. A cyclic partition that produces the best cyclic partition bound is labeled an \emph{optimal} cyclic partition, and denoted by $\Pi^*$.

For example, when $K=6$, one possible cyclic partition is $\Pi=\{\{1,3,5\}, \{4,2\},\{6\}\}$ which decomposes the users into three cycles, such that each user is represented in exactly one cycle. The corresponding cyclic partition bound is 
\begin{eqnarray}
\sum_{k=1}^6d_k&\leq&\sum_{k=1}^6\alpha_{kk}-(\alpha_{13}+\alpha_{35}+\alpha_{51})-(\alpha_{42}+\alpha_{24})-(0)
\end{eqnarray}

{\bf Participating Edge:} Edge $e_{ij}$ is a participating edge for the cyclic partition $\Pi$ if $i\neq j$ and $e_{ij}\in\pi$ for some $\pi\in\Pi$.

{\bf Cyclic Predecessor:} Under cyclic partition $\Pi$, the cyclic predecessor for user $k$ is user $\Pi(k)$, if $e_{\Pi(k) k}$ is a participating edge for $\Pi$. Note that if user $k$ belongs to a trivial cycle in $\Pi$ then $\Pi(k)=\phi$.

Finally, $\mathbb{R}^K_+$ is the set of all  $K$-tuples over non-negative real numbers.

\section{Optimality of TIN through the ADT Deterministic Model}

We first review Geng et al.'s result\footnote{Complex channel model is considered in \cite{Geng_TIN}, but the results therein are easily extended to real channel setting. Here we state the result for real channel model.} on the optimality of TIN for the $K$ user interference network with one sub-channel, i.e., $M=1$. The sub-channel index superscript is omitted in this section for compactness. 


\begin{theorem}\label{theorem:Geng}
(Theorem 1 in \cite{Geng_TIN})
In a $K$ user interference network, where the channel strength level from Transmitter $i$ to Receiver $j$ is equal to $\alpha_{ji}$, $\forall i,j\in [K]$, if the following condition is satisfied
\begin{equation}
\alpha_{ii}\geq \max_{j:j\neq i}\{\alpha_{ji}\}+\max_{k:k\neq i}\{\alpha_{ik}\},~~~\forall i,j,k\in [K],\label{eq:cond}
\end{equation}
then power control and treating interference as noise  achieve the entire GDoF region. Moreover, the GDoF region is given by

\begin{eqnarray}
\mathcal{D}_{\mbox{\tiny TIN}}&=&\left\{(d_1, d_2,\cdots, d_K)\in\mathbb{R}^K_+: \sum_{V_k\in\pi} d_k\leq \sum_{V_k\in\pi}\alpha_{kk}-w(\pi),~~\forall \pi\in[\Pi] \right\}
\end{eqnarray}

\end{theorem}

{\it Remark: Henceforth, we refer to (\ref{eq:cond}) as the TIN optimality condition for Gaussian networks. If a network (sub-channel) satisfies the TIN optimality condition (\ref{eq:cond}), the network (sub-channel) will  be referred to as a TIN optimal network (sub-channel)}.

Note that each of the bounds defining the GDoF region represents the sum-GDoF of  a cyclic interference sub-network contained in the $K$ user fully connected interference network. A cyclic sub-network is comprised of a cyclically ordered subset of users where each user causes interference only to the preceding user and suffers interference only from the following user in the cycle. As shown by Zhou et al. \cite{Zhou_Yu} and translated into the GDoF setting by Geng et al. in \cite{Geng_TIN}, the sum-GDoF of a cyclic interference sub-network is simply the sum of all desired link strengths minus the sum of all cross link strengths. For example,  the cycle $2\rightarrow 4\rightarrow 1\rightarrow 3\rightarrow 2$  corresponds to a 4 user cyclic interference sub-network with 4 desired and 4 interfering links, and its sum-GDoF are characterized by the outer bound $d_2+d_4+d_1+d_3\leq \alpha_{22}+\alpha_{44}+\alpha_{11}+\alpha_{33}-\alpha_{24}-\alpha_{41}-\alpha_{13}-\alpha_{32}$. Note that because a subset of users of cardinality $L$ has $(L-1)!$  distinct cycles, there are a total of $(L-1)!$ sum-GDoF bounds for each cardinality-$L$ subset of users, out of which all but the tightest bound are redundant. Moreover, excluding the empty set and the singletons, there are $2^K-K-1$ subsets of users that give rise to cycle bounds, some of which may again be redundant. Nevertheless, when considered together, the cycle bounds describe the precise GDoF region of the fully connected network whenever condition (\ref{eq:cond}) is satisfied. This remarkable aspect of Geng et al.'s result  greatly simplifies the proof of the outer bound of the GDoF region, because only cyclic interference networks need to be considered.

Following similar arguments as Geng et al., it is not difficult to obtain a corresponding TIN optimality result for the ADT deterministic model.
\begin{theorem} \label{single}
In a $K$ user ADT deterministic interference network, where the channel strength level from Transmitter $i$ to Receiver $j$ is equal to $n_{ji}$, $\forall i,j\in [K]$, if the following condition is satisfied
\begin{equation}
n_{ii} \geq \max_{j:j\neq i}\{n_{ji} \}+\max_{k:k\neq i}\{n_{ik} \},~~~\forall i,j,k\in [K], \label{tin}
\end{equation}
then power control and treating interference as noise can achieve the whole capacity region. Moreover, the capacity region is given by
\begin{eqnarray}
\mathcal{C}_{\mbox{\tiny TIN}}&=&\left\{(R_1, R_2,\cdots, R_K)\in\mathbb{R}^K_+: \sum_{V_k\in\pi} R_k\leq \sum_{V_k\in\pi}n_{kk}-w(\pi),~~\forall \pi\in[\Pi] \right\}
\end{eqnarray}

\end{theorem}

{\it Remark: Following a similar convention as the Gaussian case, we  refer to (\ref{tin}) as the TIN optimality condition for the ADT deterministic model. A network (sub-channel) is called TIN optimal if the TIN optimality condition (\ref{tin}) is satisfied over the network (sub-channel).}

\begin{figure}[t]
\center
\includegraphics[width=3.5 in]{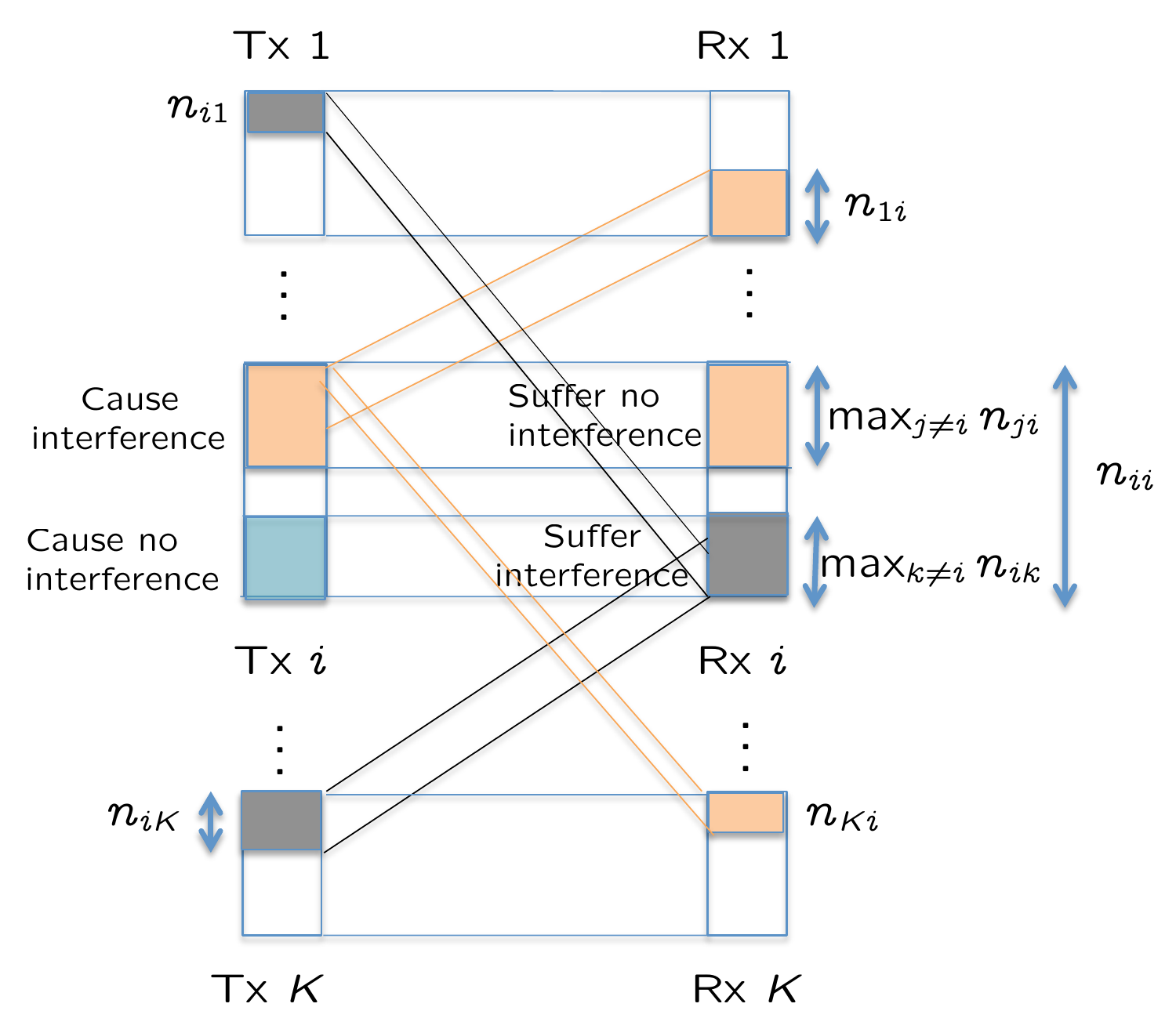} 
\caption{\small The TIN optimality condition for a $K$ user \emph{fully} connected ADT deterministic interference network.  Signal levels that cause interference do not suffer interference, and those that suffer interference cause no interference. Note that each user $i$ has $n_{ii}- \max_{j:j\neq i}\{n_{ji}\}-\max_{k:k\neq i}\{n_{ik}\}$ signal levels that neither cause interference, nor suffer interference. To avoid cluttering the figure, not all channels are shown.
}
\label{fig:cond}
\end{figure}

Note the translation from Theorem \ref{theorem:Geng} for the Gaussian case to Theorem \ref{single} for the ADT deterministic model is remarkably direct. The capacity region of the TIN optimal ADT deterministic interference network is exactly the scaled version of the GDoF region of the corresponding TIN optimal Gaussian interference network. The ADT deterministic model also reveals an interesting interpretation of the TIN optimality condition (\ref{tin}), and by association (\ref{eq:cond}). As highlighted in Figure \ref{fig:cond}, the TIN optimality condition is \emph{equivalent} to the following statements.
\begin{itemize}
\item Signal levels that suffer interference at their desired receiver, do not cause interference to others.
\item Signal levels that cause interference to others, do not suffer interference at their desired receiver.
\end{itemize}

While we omit the proof details for Theorem \ref{single} because they parallel those for Theorem \ref{theorem:Geng} presented by Geng et al. in \cite{Geng_TIN},   we will briefly present a simple alternative proof for the cycle bounds due to their central importance to this work. 

\begin{figure}[t]
\center
\includegraphics[width=2.5 in]{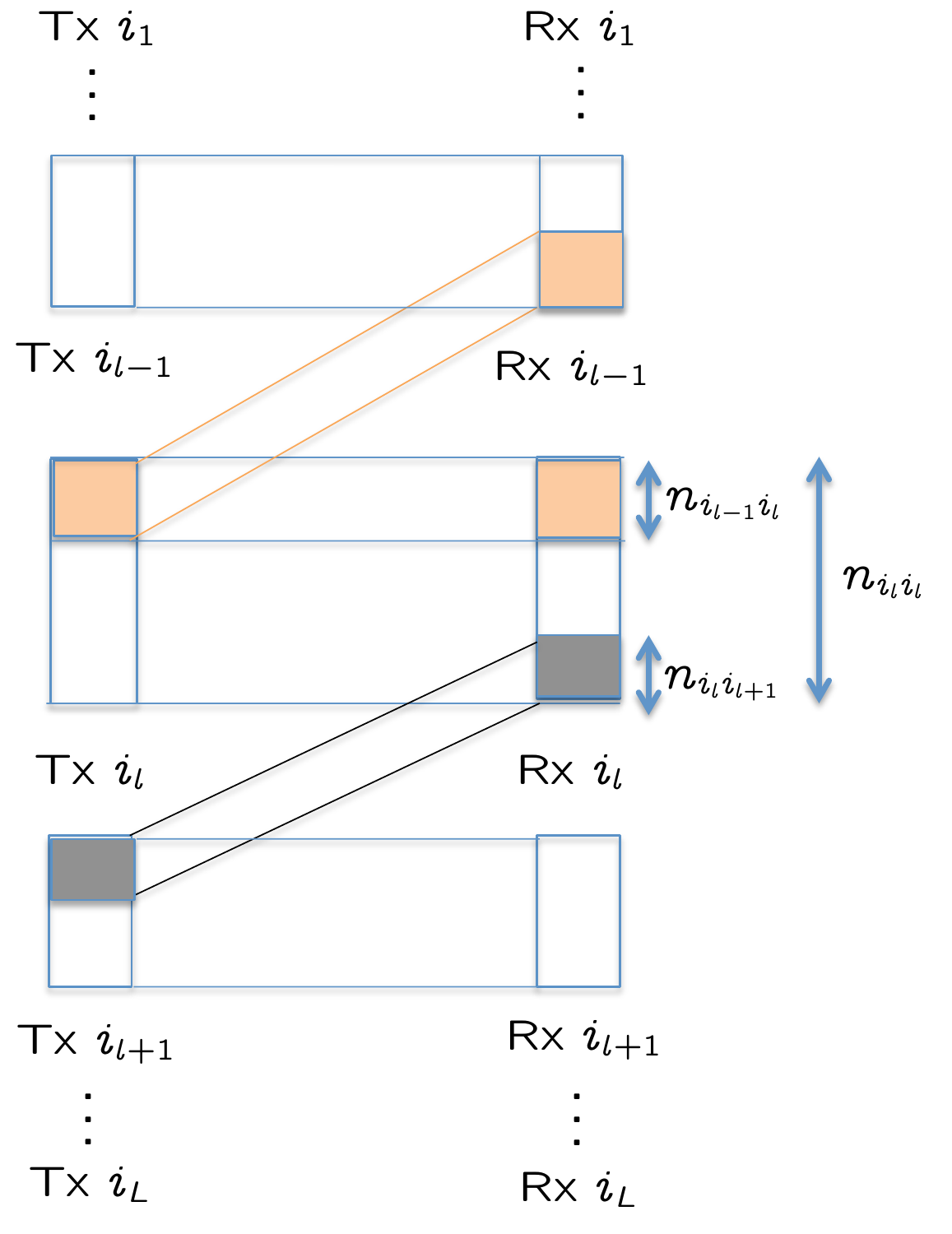} 
\caption{\small A cyclic ADT deterministic interference network that satisfies (\ref{tin}).}
\label{newcyc}
\end{figure}

Consider the cyclic interference sub-network comprised of cyclically ordered user indices $\pi = \{i_0, i_1, \ldots, i_L\}$, obtained by eliminating all remaining links, users and messages. To each receiver $i_l$, let us give all messages except $W_{i_l}, W_{i_{l+1}}$, i.e., $\{W_1,W_2,\ldots,W_K\}/ \{W_{i_l}, W_{i_{l+1}}\}$, denoted as $W_{i_l,i_{l+1}}^c$, through a genie. 
From Fano's inequality, we have
\begin{eqnarray*}
n(R_{i_l} - \epsilon) &\leq& I(W_{i_l}; Y_{i_l}^n | W_{i_l,i_{l+1}}^c) \\
&=& H(Y_{i_l}^n | W_{i_l,i_{l+1}}^c) - H(Y_{i_l}^n | W_{i_l,i_{l+1}}^c, W_{i_l}) \\
&=& H( \lfloor 2^{n_{i_l i_l}} {X}_{i_l}^n \rfloor \oplus \lfloor 2^{n_{i_l i_{l+1}}} {X}_{i_{l+1}}^n \rfloor) - H(\lfloor 2^{n_{i_l i_{l+1}}} {X}_{i_{l+1}}^n \rfloor) \\
&\overset{(a)}{=}& H(\lfloor 2^{n_{i_{l-1} i_{l}}} {X}_{i_{l}}^n \rfloor) + H(\lfloor 2^{n_{i_l i_l}} {X}_{i_l}^n \rfloor \oplus \lfloor 2^{n_{i_l i_{l+1}}} {X}_{i_{l+1}}^n \rfloor | \lfloor 2^{n_{i_{l-1} i_{l}}} {X}_{i_{l}}^n \rfloor) - H(\lfloor 2^{n_{i_l i_{l+1}}} {X}_{i_{l+1}}^n \rfloor) \\
&\overset{(b)}{\leq}& n( n_{i_l i_l} - n_{i_{l-1} i_l} ) + H(\lfloor 2^{n_{i_{l-1} i_{l}}} {X}_{i_{l}}^n \rfloor)  - H(\lfloor 2^{n_{i_l i_{l+1}}} {X}_{i_{l+1}}^n \rfloor)
\end{eqnarray*}
where $(a)$ follows from the assumption $n_{i_l i_l} \geq n_{i_{l-1} i_l} + n_{i_l i_{l+1}}$ such that the interfering-causing bits $\lfloor 2^{n_{i_{l-1} i_{l}}} {X}_{i_{l}}^n \rfloor$ suffer no interference at the desired receiver $i_l$ and $(b)$ is due to the fact that the entropy of a variable is no more than the number of bits therein. See Figure \ref{newcyc} for a pictorial illustration.
Adding the above inequalities for $l \in \{1,2,\ldots,L\}$, we find that the entropy terms cancel out  leaving us with
\begin{eqnarray*}
\sum_{l=1}^L n(R_{i_l} - \epsilon) &\leq& \sum_{l=1}^L n( n_{i_l i_l} - n_{i_{l-1} i_l} ) = n \sum_{l=1}^L n_{i_l i_l} - n w(\pi)
\end{eqnarray*}
from which we arrive at the desired bound by normalizing by $n$ on both sides of the inequality and letting $n$ approach infinity .

{\it Remark: Henceforth, since we are only interested in networks that satisfy the TIN optimality conditions, (\ref{tin}) in the deterministic setting and (\ref{eq:cond}) in the Gaussian setting, we will assume throughout that these conditions are satisfied.}

\section{Sum-Capacity (Sum-GDoF)}
We now switch our attention from capacity region to sum-capacity in the deterministic case, and from GDoF region to sum-GDoF in the Gaussian case. To avoid repetition, we will focus the discussion in this section to the Gaussian setting, i.e., GDoF  region, sum-GDoF, channel strengths $\alpha_{ij}$, etc., but all arguments made in this section also apply to the deterministic setting, with capacity region,  sum-capacity, channel strengths $n_{ij}$. 

Since we already have the GDoF region characterization in Theorem \ref{theorem:Geng}, the sum-GDoF characterization may appear trivial. However, there are certain interesting aspects of this problem that we will highlight in this section, which will be especially useful when we move on to parallel interference networks in subsequent sections.

Consider, for example, the  GDoF region of the TIN optimal 3 user interference network, which is the set of  tuples $(d_1, d_2, d_3)\in\mathbb{R}^3_+$, defined by the following constraints.
\allowdisplaybreaks
\begin{eqnarray}
 d_1 &\leq& \alpha_{11}-w(\{1\})=\alpha_{11}\label{r1} \\
d_2 &\leq& \alpha_{22}-w(\{2\}) =\alpha_{22}\label{r2} \\
d_3 &\leq& \alpha_{33}-w(\{3\}) =\alpha_{33}\label{r3} \\
d_1 + d_2 &\leq& \alpha_{11} + \alpha_{22} -w(\{1,2\})=\alpha_{11} + \alpha_{22} -\alpha_{12}-\alpha_{21}\label{r4} \\
d_2 + d_3 &\leq& \alpha_{22} + \alpha_{33} -w(\{2,3\})=\alpha_{22} + \alpha_{33} -\alpha_{23}-\alpha_{32}\label{r5} \\
d_3 + d_1 &\leq& \alpha_{33} + \alpha_{11} -w(\{3,1\})=\alpha_{11} + \alpha_{33} -\alpha_{31}-\alpha_{13}\label{r6} \\
d_1 + d_2 + d_3 &\leq& \alpha_{11} + \alpha_{22} + \alpha_{33} -w(\{1,2,3\})=\alpha_{11} + \alpha_{22}+\alpha_{33} -\alpha_{12}-\alpha_{23}-\alpha_{31}\label{r7} \\
d_1 + d_2 + d_3 &\leq& \alpha_{11} + \alpha_{22} + \alpha_{33} -w(\{3,2,1\})=\alpha_{11} + \alpha_{22} + \alpha_{33} -\alpha_{21} - \alpha_{32} - \alpha_{13}\label{r8}
\end{eqnarray}
The last two bounds are already sum-GDoF bounds. However, remarkably, neither of these may be tight. This is because, unlike similar forms that are commonly encountered e.g.,  the capacity region of the multiple access channel, this region is not polymatroidal. It is easy to see that a direct sum of (\ref{r1}) and (\ref{r6}), for example, \emph{could} provide a tighter sum-GDoF bound. Incidentally, this would be a cyclic partition bound for the cyclic partition $\Pi=\{\{1\},\{2,3\}\}$. But, how about something a bit more involved, such as $1/2$ times the sum of (\ref{r4}), (\ref{r5}), (\ref{r6}), which would also produce a sum-rate bound (but not a cyclic partition bound)? Let us consider this bound.
\begin{eqnarray}
\frac{(\ref{r4})+(\ref{r5})+(\ref{r6})}{2}\Rightarrow d_1+d_2+d_3&\leq &\sum_{k=1}^3\alpha_{kk}-\frac{w(\{1,2\})+w(\{2,3\})+w(\{3,1\})}{2}
\end{eqnarray}
Interestingly, this is the same bound as $1/2$ times the sum of $(\ref{r7})$ and $(\ref{r8})$. Therefore, it can never be tighter than the tightest of $(\ref{r7})$ and $(\ref{r8})$. Therefore, even though the GDoF region is not polymatroidal, the special structure of the cycle bounds imparts some special properties. This is what we will explore in this section. In fact, these examples  are representative of our general result. We will show that for a TIN optimal $K$ user interference network, the sum-GDoF value is always given by a cyclic partition bound. This is the main result of this section, and we state it in the following theorem.
\begin{theorem}\label{decompose}
For TIN optimal Gaussian interference networks
\begin{eqnarray}
\mathcal{D}_\Sigma &=&\mathcal{D}^{\Pi*}_{\Sigma}
\end{eqnarray}
where $\mathcal{D}^{\Pi*}_{\Sigma}$ is the best cyclic partition bound.
\end{theorem}
\proof 
The sum-GDoF value is expressed by the linear program
\begin{align}
(LP_1)~~~~~~~  \mathcal{D}_\Sigma=\max & \mbox{ }d_1+d_2+\cdots+d_K\\
\mbox{such that}& \sum_{V_k\in\pi}d_k\leq \sum_{V_k\in\pi}\alpha_{kk}-w(\pi), ~~\forall \pi\in[\Pi]\\
&d_k\geq 0, ~~ \forall k\in[K] \label{eq:positive}
\end{align}


In Section \ref{sec:sign} we show that the non-negativity constraint (\ref{eq:positive}) can be eliminated from $LP_1$ without affecting its value. This allows us to express the sum-GDoF in terms of the dual LP as follows.
\begin{align}
(LP_2)~~~~~~~ \mathcal{D}_\Sigma=\min~ & \sum_{\pi\in\Pi}\lambda_\pi\left( \sum_{V_k\in\pi}\alpha_{kk}-w(\pi)\right)\\
\mbox{such that}&\sum_{\pi\in\Pi}\lambda_\pi 1(V_k\in\pi)=1, ~~\forall k\in[K]\\
&\lambda_\pi\geq 0, ~~ \forall \pi\in[\Pi] 
\end{align}
where $1(\cdot)$ is the indicator function that returns the values 1 or 0 when the argument to the function is true or false, respectively.

\noindent Equivalently,
\begin{align}
(LP_3)~~~~~~~ \mathcal{D}_\Sigma=& \sum_{k=1}^K\alpha_{kk}-\max\sum_{\pi\in\Pi}\lambda_\pi w(\pi)\\
\mbox{such that}&\sum_{\pi\in\Pi}\lambda_\pi 1(V_k\in\pi)=1, ~~\forall k\in[K]\label{eq:lambdaconstraint}\\
&\lambda_\pi\geq 0, ~~ \forall \pi\in[\Pi] 
\end{align}

\noindent Let us also define the integer constrained version of this LP.
\begin{align}
(IP_4)~~~~~~~ \mathcal{D}^{\Pi*}_{\Sigma}=& \sum_{k=1}^K\alpha_{kk}-\max\sum_{\pi\in\Pi}\lambda_\pi w(\pi)\\
\mbox{such that}&\sum_{\pi\in\Pi}\lambda_\pi 1(V_k\in\pi)=1, ~~\forall k\in[K]\\
&\lambda_\pi\in\{0,1\}, ~~ \forall \pi\in[\Pi] 
\end{align}
Note that the integer program $IP_4$ is simply the best cyclic partition bound $\mathcal{D}^{\Pi*}_{\Sigma}$.

Since imposing an integer constraint cannot make the $\max$ term larger, it is already clear that $\mathcal{D}^{\Pi*}_{\Sigma}\geq\mathcal{D}_\Sigma$. To prove the other direction, let us reformulate $LP_3$ by changing the perspective from cycles to edges. Instead of the multipliers $\lambda_\pi$ that are associated with cycles, we will use multipliers $t_{ij}$ that are associated with edges. Define
\begin{eqnarray}
t_{ij}&=&\sum_{\pi\in\Pi} \lambda_\pi 1(e_{ij} \in\pi), ~~~\forall (i,j)\in[K]\times[K] \label{t}\label{eq:tconstraint}
\end{eqnarray}

We  now translate the constraints (\ref{eq:lambdaconstraint}) on cycles to edges. A cycle incident on vertex $k$ must have exactly one incoming and one outgoing edge. (\ref{eq:lambdaconstraint}) says that the net contribution from $\lambda_\pi$ for all cycles associated with any particular vertex is 1. Clearly, then  the net contribution for all edges leaving a transmitter (vertex), or all edges entering a receiver (vertex), must be unity.
\begin{eqnarray}
\sum_{j=1}^Kt_{ij}&=&1, ~\forall i\in[K]\\
\sum_{i=1}^Kt_{ij}&=&1, ~\forall j\in[K]
\end{eqnarray}
and the objective value is equivalently re-written as
\begin{eqnarray}
\sum_{\pi\in\Pi} \lambda_\pi w(\pi) &=&\sum_{\pi\in\Pi} \lambda_\pi \sum_{e_{ij}\in\pi}w(e_{ij})\\
&=&\sum_{\pi\in\Pi} \lambda_\pi \sum_{(i,j)\in[K]\times[K]}w(e_{ij})1(e_{ij}\in\pi)\\
&=&\sum_{(i,j)\in[K]\times[K]}w(e_{ij})\sum_{\pi\in\Pi} \lambda_\pi 1(e_{ij}\in\pi)\\
&=&\sum_{(i,j)\in[K]\times[K]}t_{ij}w(e_{ij})
\end{eqnarray}

\noindent Substituting into $LP_3$, this gives us the new LP
\begin{align}
(LP_5)~~~~~~~ \mathcal{D}_\Sigma\geq& \sum_{k=1}^K\alpha_{kk}+\min\sum_{(i,j)\in[K]\times[K]}c_{ij}t_{ij}\\
\mbox{such that }&\sum_{j=1}^Kt_{ij}=1, ~\forall i\in[K]\\
&\sum_{i=1}^Kt_{ij}=1, ~\forall j\in[K]\\
&t_{ij}\geq 0, ~\forall (i,j)\in[K]\times[K]
\end{align}
where we defined $c_{ij}=-w(e_{ij})$, and the $\geq$ sign appears because we dropped the constraint (\ref{eq:tconstraint}).  In this standard form, this LP is recognizable as the \emph{minimum weight perfect matching} problem, and its solution is known to be integral, i.e., the optimizing $t_{ij}$ must take values in $\{0,1\}$ (See \cite{Schrijver}   and Theorem 5 of \cite{matching}).

However, note that any integral  solution to $LP_5$ gives us a valid cyclic partition bound, $\mathcal{D}^{\Pi}_\Sigma$. Therefore we have,
\begin{eqnarray}
\mathcal{D}_\Sigma&\geq&\mathcal{D}^{\Pi}_\Sigma\\
&\geq&\mathcal{D}^{\Pi*}_\Sigma
\end{eqnarray}
because a cyclic partition bound cannot be smaller than the optimal cyclic partition bound. Since we have already shown that $\mathcal{D}_\Sigma\leq \mathcal{D}^{\Pi*}_\Sigma$, we must have $\mathcal{D}^{\Pi*}_\Sigma=\mathcal{D}_\Sigma$. \hfill\QED

Finally, since the same proof also works for the deterministic setting, let us conclude this section by stating the deterministic counterpart of Theorem \ref{decompose} as the following corollary.
\begin{corollary}\label{col:decompose}
For TIN optimal ADT deterministic interference networks
\begin{eqnarray}
\mathcal{C}_\Sigma &=&\mathcal{C}^{\Pi*}_{\Sigma}
\end{eqnarray}
where $\mathcal{C}^{\Pi*}_{\Sigma}$ is the best cyclic partition bound.
\end{corollary}

\section{Optimality of TIN for Parallel Interference Networks}
As we move from the single sub-channel case to multiple parallel sub-channels, the outer bound proof becomes significantly more challenging. Whereas formerly it was sufficient to only consider each cyclic sub-network obtained by eliminating all other users, messages and links, this is no longer possible for parallel interference networks. For example, a different cycle may be active in each sub-channel, however one cannot eliminate a different set of links for each sub-channel. As an outer bounding argument, eliminating a link is justified by including a genie that takes all the messages originating at the transmitter of that link, and provides them to the receiver of that link, so that the receiver can reconstruct and subtract the transmitted symbols from its received signal. However, in a parallel channels setting, the message information provided by the genie allows a receiver to reconstruct and subtract the transmitted symbols from a transmitter on \emph{all} sub-channels. Thus, if a link from Transmitter $i$ to Receiver $j$ is removed for one sub-channel, it must be removed for all sub-channels. This makes it impossible to reduce a fully connected parallel interference network directly into \emph{different} cyclic sub-networks over each sub-channel. As such, for parallel interference networks, the reduction to cyclic networks is in general no longer an option, and the entire network must be directly considered for the outer bound.
Given this added source of difficulty, the relative simplicity of the ADT deterministic model is tremendously useful. Thus, we start to explore  parallel interference networks with the ADT deterministic model.

\subsection{ADT Deterministic Model}
While we deal with multiple parallel sub-channels in this section, recall that we  assume throughout that each sub-channel satisfies condition (\ref{tin}). In other words, by itself, each sub-channel is TIN optimal. What we wish to explore is whether  \emph{collectively} such parallel channels remain separable and therefore TIN optimal. Let us start with a few relevant definitions. 

For the definitions that have been introduced for the single sub-channel case, we will add a superscript to indicate the sub-channel index, for example cyclic partition $\Pi^{[m]}$, cyclic predecessor $\Pi^{[m]}(k)$,  and cyclic partition bound $\mathcal{C}_{\Sigma}^{\Pi^{[m]}}$. Note that many cyclic partitions are possible for each sub-channel, and a different cyclic partition may be used for each sub-channel.

{\bf Participating Input and Output Levels ($X_{i,u}^{[m]}, Y_{k,u}^{[m]}$)}: For the $m$-th sub-channel, we define participating input levels  $$X_{i,u}^{[m]} \triangleq 0.X_{i,(1)}^{[m]}, \ldots, X_{i,\left(n_{\Pi^{[m]}{(i)} i}^{[m]}\right)}^{[m]}$$ to be the bits that are sent from Transmitter $i$ and observed at its predecessor Receiver $\Pi^{[m]} (i)$. The received signal levels resulting from all interfering $X_{i,u}^{[m]}$ are defined as the participating output levels $$Y_{k,u}^{[m]} \triangleq \sum_{i=1,i \neq k}^K 2^{n_{ki}^{[m]}} {X}_{i,u}^{[m]}$$ where the summation is bit-wise modulo two. We can also write $X_{i,u}$ in a vector form as $$X_{i,u}^{[m]} = [ X_{i,(1)}^{[m]}, \ldots, X_{i,\left(n_{\Pi^{[m]}(i) i}^{[m]}\right)}^{[m]} ].$$ Similar vector notation is used for $Y_{k,u}^{[m]}$ when the vector form is clearer. 

{\bf Invertibility}: The $m$-th sub-channel is said to be invertible  if the  mapping from ${\bf X}_u^{[m]} \triangleq (X_{1,u}^{[m]}, \ldots, X_{K,u}^{[m]})$ to ${\bf Y}_u^{[m]} \triangleq (Y_{1,u}^{[m]}, \ldots, Y_{K,u}^{[m]})$ is invertible for an optimal cyclic partition $\Pi^{[m]*}$. Mathematically, we require 
\begin{eqnarray}
H({\bf X}_u^{[m]}|{\bf Y}_u^{[m]})&=&0.\label{eq:invertADT}
\end{eqnarray}





The significance of these definitions will become clear with the statement of the result, illustrative examples, and finally from the details of the  proof. Perhaps the most intriguing is the invertibility property. At this point it suffices to say that it is a ``mild" property and is easily testable for a given problem instance. The mildness of this property will be  explicitly addressed in Section \ref{sec:mild}.
With these definitions, we are now ready to state the main result of this section in the following theorem.

\begin{theorem}\label{Kdetthm}
In a $K$ user ADT deterministic interference network with $M$ sub-channels, 
if each sub-channel is individually  TIN optimal and invertible, then even collectively for all the sub-channels of the parallel interference network, the sum-capacity is achieved by a separate TIN solution over each sub-channel. 
\end{theorem}

The proof of Theorem \ref{Kdetthm} is deferred to Section \ref{p1}. At this point it is  important to understand the statement of the theorem and its limitations through illustrative examples.

\begin{figure}[h]
\center
\includegraphics[width=6in]{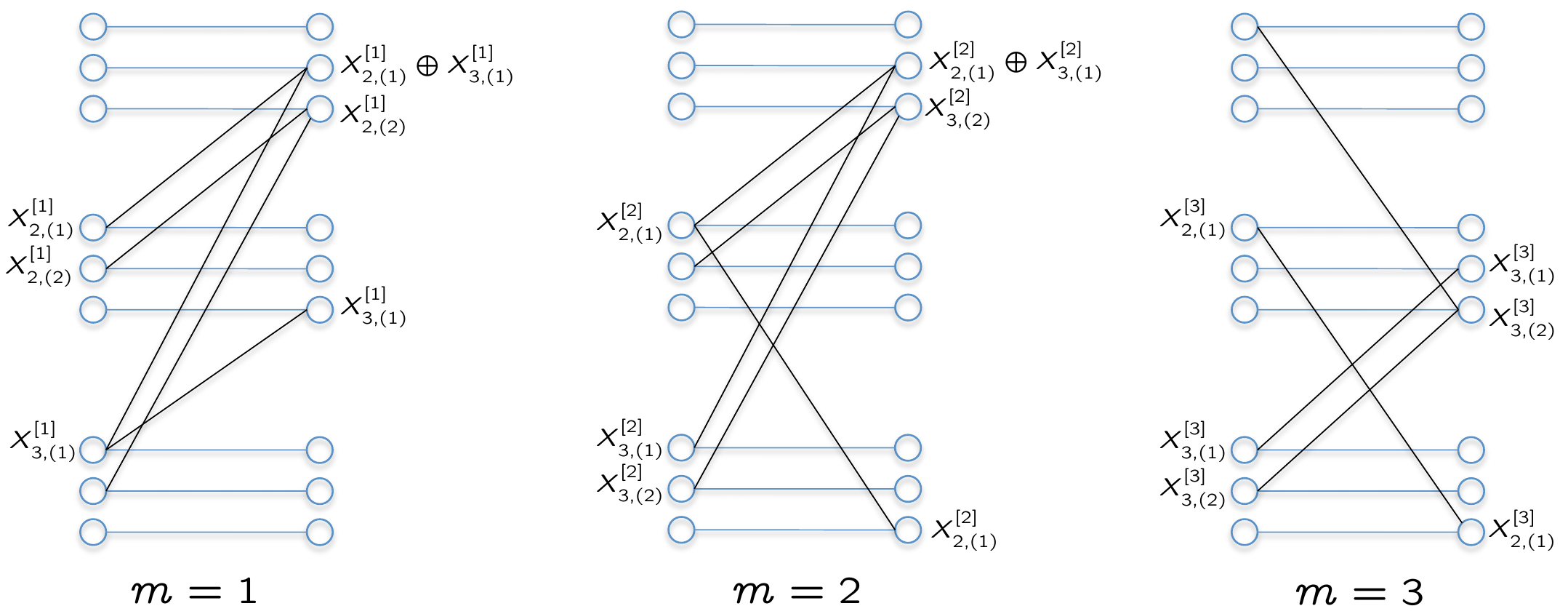}
\caption{\small A 3 user ADT deterministic interference network with 3 sub-channels, where each sub-channel is TIN optimal. Under the optimal cyclic partitions $\Pi^{[1]*}=\{\{1,2,3\}\}, \Pi^{[2]*}=\{\{3,2,1\}\}, \Pi^{[3]*}=\{\{1\}, \{2,3\}\}$, the participating input and output levels, $X_{i,u}^{[m]}, Y_{i,u}^{[m]}, i,m \in \{1,2,3\}$ are labeled and  the mapping from $(X_{1,u}^{[m]}, X_{2,u}^{[m]}, X_{3,u}^{[m]})$ to $(Y_{1,u}^{[m]}, Y_{2,u}^{[m]}, Y_{3,u}^{[m]})$ is easily verified to be invertible for each sub-channel.}
\label{ex1}
\end{figure}

\begin{example}
Consider the $K=3$ user ADT deterministic interference network with $M=3$ parallel sub-channels, shown in Figure \ref{ex1}. It is readily verified that each sub-channel by itself is TIN optimal. For example, consider user 2 in sub-channel 1. The desired signal strength for this user is $n_{22}^{[1]}=3$, the strongest interference caused by this user is $n_{12}^{[1]}=2$ and the strongest interference suffered by this user is $n_{23}^{[1]}=1$. Thus, the desired signal strength is no less than the sum of the signal strengths of the strongest interference caused and the strongest interference received by this user. The same is true for each of the 3 users in each of the 3 parallel sub-channels. Therefore, according to Theorem \ref{single}, TIN is optimal for each sub-channel by itself.
For the 3 sub-channels, consider the optimal cyclic partitions $$\Pi^{[1]*}=\{\{1,2,3\}\}, \Pi^{[2]*}=\{\{3,2,1\}\}, \Pi^{[3]*}=\{\{1\}, \{2,3\}\}.$$
The weights of the participating edges are 
\begin{align}
w(\Pi^{[1]*}) &= w(\{e_{12}^{[1]}, e_{23}^{[1]}, e_{31}^{[1]} \}) = n_{12}^{[1]} + n_{23}^{[1]} + n_{31}^{[1]}  = 3 \\
w(\Pi^{[2]*}) &= w(\{e_{32}^{[2]}, e_{21}^{[2]}, e_{13}^{[2]} \}) = n_{32}^{[2]} + n_{21}^{[2]} + n_{13}^{[2]}  = 3 \\
w(\Pi^{[3]*}) &= w(\{e_{11}^{[3]}, e_{23}^{[3]}, e_{32}^{[3]} \}) = 0 + n_{23}^{[3]} + n_{32}^{[3]}  = 3
\end{align}
Then according to Corollary \ref{col:decompose}, the sum-capacity values for each sub-channel by itself are given by $$\mathcal{C}^{[m]}_{\Sigma}=  \sum_{i=1}^3 n_{ii}^{[m]} - w(\Pi^{[m]*}) =9-3=6, ~m = 1,2,3.$$ What we wish to know is if TIN continues to be the sum-capacity optimal scheme for all 3 sub-channels collectively. 

Let us check for invertibility for each sub-channel. According to the definitions, the participating inputs for sub-channel 1 are $X_{1,u}^{[1]} = [X_{1,(1)}^{[1]}, \ldots, X_{1,(n_{31}^{[1]})}^{[1]}] = \phi, X_{2,u}^{[1]} = [X_{2,(1)}^{[1]}, \ldots, X_{2,(n_{12}^{[1]})}^{[1]}] = [ X_{2,(1)}^{[1]}, X_{2,(2)}^{[1]} ]$, $X_{3,u}^{[1]} = [X_{3,(1)}^{[1]}, \ldots, X_{3,(n_{23}^{[1]})}^{[1]}] = [ X_{3,(1)}^{[1]}]$ and the participating outputs for sub-channel 1 are $Y_{1,u}^{[1]} = [X_{2,(1)}^{[1]} \oplus X_{3,(1)}^{[1]}, X_{2,(2)}^{[1]}], Y_{2,u}^{[1]} = [X_{3,(1)}^{[1]}]$ and $Y_{3,u}^{[1]} = \phi$. It is now trivial to verify that from $(Y_{1,u}^{[1]}, Y_{2,u}^{[1]}, Y_{3,u}^{[1]})$, we can recover $(X_{1,u}^{[1]},X_{2,u}^{[1]},X_{3,u}^{[1]})$. Therefore, sub-channel 1 is invertible. Similarly, the participating inputs and outputs for sub-channels 2 and 3 are shown in Figure \ref{ex1} and it is  easily verified that sub-channels 2 and 3 are invertible as well. Therefore, since all the conditions of Theorem \ref{Kdetthm} are satisfied, we conclude that separate TIN is optimal for this parallel interference network, and therefore, the sum-capacity of the 3 sub-channels collectively, is the sum of their individual sum-capacities. In other words, the sum-capacity is $6+6+6=18$ and is achieved by separate TIN on each sub-channel.

\end{example}

To also expose the limitation of Theorem \ref{Kdetthm}, the next example illustrates a relatively rare situation where invertibility is not satisfied, and so  Theorem \ref{Kdetthm} cannot be applied.

\begin{figure}[h]
\center
\includegraphics[width=6in]{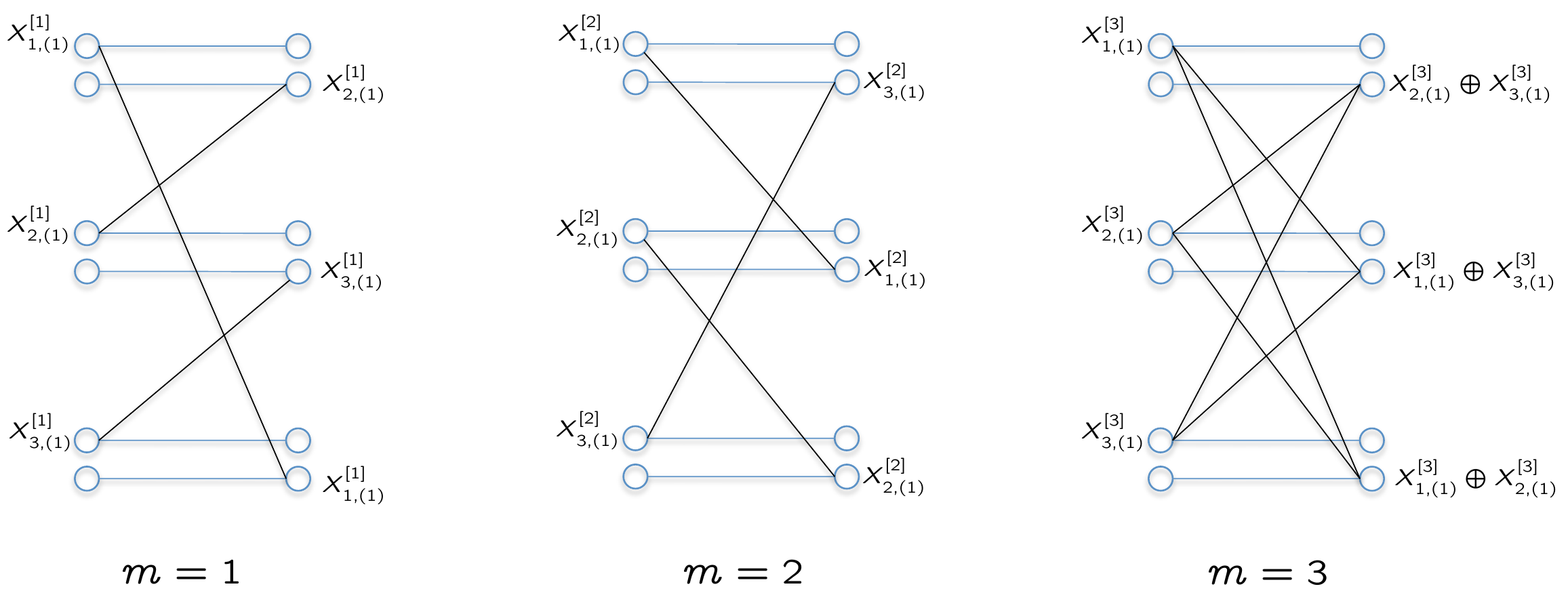}
\caption{\small A 3 user ADT deterministic interference network with 3 sub-channels, where each sub-channel is TIN optimal. For the optimal cyclic partitions  $\Pi^{[1]*}=\{ \{ 1,2,3 \} \}, \Pi^{[2]*}=\{ \{ 3,2,1\}\}$ and $\Pi^{[3]*}=\{ \{ 1,2,3 \} \}$, participating inputs and outputs $X_{i,u}^{[m]}, Y_{i,u}^{[m]}, i, m \in \{1,2,3\}$ are labeled. In this case, the mapping from $(X_{1,u}^{[3]}, X_{2,u}^{[3]}, X_{3,u}^{[3]})$ to $(Y_{1,u}^{[3]}, Y_{2,u}^{[3]}, Y_{3,u}^{[3]})$ is not invertible.}
\label{ex2}
\end{figure}
\begin{example}
Consider the 3 user ADT deterministic interference network with 3 sub-channels, as shown in Figure \ref{ex2}, with the optimal cyclic partitions  $\Pi^{[1]*}=\{ \{ 1,2,3 \} \}, \Pi^{[2]*}=\{ \{ 3,2,1\}\}$ and $\Pi^{[3]*}=\{ \{ 1,2,3 \} \}$ for the first, second and third sub-channel, respectively. It is easy to verify that all 3 sub-channels are TIN optimal individually. However, with the participating inputs and outputs $X_{i,u}^{[m]}, Y_{i,u}^{[m]}$ shown in the figure, it is also easy to see while the first two sub-channels are invertible, the third sub-channel is not.
\end{example}

Note that when the network only has one sub-channel, i.e., $M=1$, we can delete all the interfering links except the participating interference links (ones in $\Pi^{[1]}$) without violating the outer bound argument, so that the invertibility becomes trivially true. Thus, Theorem \ref{Kdetthm} recovers the outer bound result of Theorem \ref{single}. 

There are many interesting classes of networks where invertibility is shown to hold easily. For example, when $K=3$, then invertibility is fully characterized in Section \ref{sec:mild}. Another interesting class is the class of cyclic interference networks where each sub-channel contains only one cycle  (different sub-channels may have different cycles). These and other interesting cases will be discussed in Section \ref{sec:mild}.

\subsection{GDoF}

We now explore the extension to the Gaussian setting and show that the insights from the deterministic framework go through. We obtain the corresponding result on  the sum-GDoF optimality of TIN for parallel Gaussian interference networks subject to similar invertibility property. $X_{i,u}^{[m]}, Y_{k,u}^{[m]}$ are defined similar to the deterministic case. Participating input bit levels $X_{i,u}^{[m]}$ are  made up of the bit levels below the decimal point, sent from Transmitter $i$ and heard by Receiver $\Pi^{[m]}(i)$, i.e., $X_{i,u}^{[m]} = \mbox{sign}(X_i^{[m]}) \times 0.X_{i,(1)}^{[m]},\ldots,X_{i,\left(n_{\Pi^{[m]}(i)i}^{[m]}\right)}^{[m]}$, where $n_{ki}^{[m]} = \lfloor \frac{1}{2} \alpha_{ki}^{[m]} \log_2 P \rfloor$. Participating output levels $Y_{k,u}^{[m]}$ are the resulting interference from $X_{i,u}^{[m]}$ plus additive Gaussian noise, i.e.,  $Y_{k,u}^{[m]} = \sum_{i=1,i \neq k}^K h_{ki}^{[m]} X_{i,u}^{[m]} + Z_k^{[m]}$. 

 
The invertibility property is a bit more delicate to translate, because of the presence of noise, average power constraints, and the focus on GDoF rather than exact capacity. Given a cyclic partition, for the invertibility property in the Gaussian case, it suffices to require the mapping from ${\bf X}_u^{[m]} \triangleq (X_{1,u}^{[m]}, \ldots, X_{K,u}^{[m]})$ to ${\bf Y}_u^{[m]} \triangleq (Y_{1,u}^{[m]}, \ldots, Y_{K,u}^{[m]})$ to be invertible \emph{within bounded noise distortion}. 
Mathematically we express the counterpart of (\ref{eq:invertADT})  as
\begin{eqnarray}
\mbox{\bf (Invertibility Property): } H({\bf X}_u^{[m]} | {\bf Y}_u^{[m]})&=& o(\log(P)) \label{equ:ginv}
\end{eqnarray}
As before, the $m$-{th} sub-channel is said to be  invertible if there exists an optimal cyclic partition $\Pi^{[m]*}$ under which invertibility is satisfied.

We have the following theorem.

\begin{theorem}\label{thm}
In a $K$ user parallel Gaussian interference network with $M$ sub-channels, 
if each sub-channel is  individually both TIN optimal and invertible, then the sum-GDoF value of the parallel Gaussian interference network is achieved by separate TIN over each sub-channel.
\end{theorem}

The proof of Theorem \ref{thm} appears in Section \ref{p3}. 

\subsection{Mildness of Invertibility Condition}\label{sec:mild}
The intuition behind the mildness of the invertibility condition is analogous to the commonly encountered issue of invertibility of channel matrices in wireless networks, i.e., the property is satisfied everywhere except over an algebraic variety of lower dimension than the parameter space, and therefore is increasingly likely to be true when the parameter space is a large field. In particular, we expect  invertibility to hold in the Gaussian setting almost surely. In the deterministic setting also, because the signal levels $n_{ij}$ are defined as quantized versions of $\alpha_{ij}\log(P)$, with $\alpha_{ij}$ drawn from a continuum of real values, as the quality of the quantization improves (with increasing $P$), the invertibility is increasingly likely to hold.

To  strengthen this intuition, we take a closer look at the invertibility condition in this section. We will go into details mainly for the deterministic setting. For the Gaussian setting, while the insights from deterministic setting are expected to go through via the usual machinery of translating between deterministic and Gaussian settings, as used in a number of works \cite{Bresler_Tse, Bresler_Parekh_Tse, Avestimehr_Diggavi_Tse, Avestimehr_Sezgin_Tse, Jafar_Vishwanath_GDOF, Niesen_Maddah_Ali_X}, an in-depth analysis appears to be extremely cumbersome with little by way of new insights. Hence we will restrict the discussion in the Gaussian setting primarily to just an intuitive level.

\subsubsection{ADT Deterministic Model}
\paragraph{3 users}
 Let us start with the ADT deterministic model for $K=3$, with arbitrary $M$, where we explicitly characterize the invertibility condition.
\begin{lemma}\label{lemma:3inv}
For the $m$-{th} sub-channel of a 3 user ADT deterministic interference network, if $n_{12}^{[m]} + n_{23}^{[m]} + n_{31}^{[m]} \neq n_{21}^{[m]} + n_{32}^{[m]} + n_{13}^{[m]}$, then sub-channel $m$ is invertible under any  cyclic partition.
\end{lemma}
{\it Proof:} Consider the bi-partite graph comprised of the participating input and output levels as the two sets of vertices and the cross links  between them as the edges. According to Theorem  \ref{lemma:inv}, if this graph is acyclic then invertibility must hold. Therefore, we only need to show that when $n_{12}^{[m]} + n_{23}^{[m]} + n_{31}^{[m]} \neq n_{21}^{[m]} + n_{32}^{[m]} + n_{13}^{[m]}$, the bipartite graph is acyclic. Let us suppose the opposite, i.e., the graph has a cycle. Since only cross links are considered, for the 3 user case, the cycle must  must traverse all 3 users. The 6 edges along the way correspond to 6 interfering links with strength $n_{ji}^{[m]}$. The bit sent from Transmitter $i$ to Receiver $j$ is shifted $n_{ji}^{[m]}$ places. Therefore as we traverse the 6 edges, the net shift factor encountered is  $n_{12}^{[m]} + n_{23}^{[m]} + n_{31}^{[m]} - n_{21}^{[m]} - n_{32}^{[m]} - n_{13}^{[m]}$, which must equal zero for the cyclical path to return to its origin. But this contradicts the assumption that $n_{12}^{[m]} + n_{23}^{[m]} + n_{31}^{[m]} \neq n_{21}^{[m]} + n_{32}^{[m]} + n_{13}^{[m]}$. This completes the proof by contradiction.
\hfill\QED

Combining the result of Lemma \ref{lemma:3inv} with the result of Theorem \ref{Kdetthm}, we have the explicit result for the 3 user parallel ADT deterministic interference network.

\begin{theorem}\label{3detthm}
For the $3$ user parallel ADT deterministic interference network where each sub-channel is individually TIN optimal, if each sub-channel also satisfies
\begin{eqnarray}
n_{12}^{[m]} + n_{23}^{[m]} + n_{31}^{[m]} \neq n_{21}^{[m]} + n_{32}^{[m]} + n_{13}^{[m]}, \forall m \in[M] \label{detrd}
\end{eqnarray}
then the sum-capacity of the $3$ user parallel ADT deterministic interference network is achieved by a separate TIN solution over each sub-channel.
\end{theorem}

\paragraph{Acyclic Bipartite Graph of Cross Channels between Participating Levels (Includes Cyclic Interference Networks)}

The following theorem  presents a general result  which was also used in the proof of invertibility for the 3 user case.

\begin{theorem}\label{lemma:inv}
For each sub-channel of a $K$ user parallel ADT deterministic  interference network, view the cross links between the participating input and output levels as the edges of an undirected bipartite graph. If this bipartite graph is acyclic, then the sub-channel is invertible. If each sub-channel individually is TIN optimal, then separate TIN over each sub-channel achieves the sum-capacity of the $K$ user parallel ADT deterministic  interference network.
\end{theorem}

{\it Proof:}  Since the optimality of separate TIN is already established subject to invertibility, all that remains is to show that invertibility holds. We will prove that in the absence of cycles in the bi-partite graph described above, one can always start from any participating input bit level as the root and build a tree with participating output bit levels as leaves such that we can proceed to the end of the tree (leaves) and start inverting sequentially from participating output levels to recover all participating input levels along the tree. The construction is as follows. Start at any participating input bit level as the root. When we leave the input bit level for an output bit level, always choose a \emph{participating} edge. 
Note that for each input bit level, there is only one participating edge. Also, there is only one participating edge for each output bit level. After reaching the output bit level, if it is connected nowhere else then this is the leaf and we are done. If it is connected to other input bit levels, the edges must all be \emph{non-participating} edges as the only participating edge has been used to arrive at the output bit. Again, for each  input level reached, choose the only participating edge to reach the next output bit level. Because the graph has no cycles, the process must end eventually. We cannot end at an input level, because every input bit level must have a participating edge going out.  Therefore we must end at output bit levels (leaves). Then we can traverse this tree back and find the original input bit level and all input bits along the way. \hfill\QED

\begin{figure}[h]
\center
\includegraphics[width=5 in]{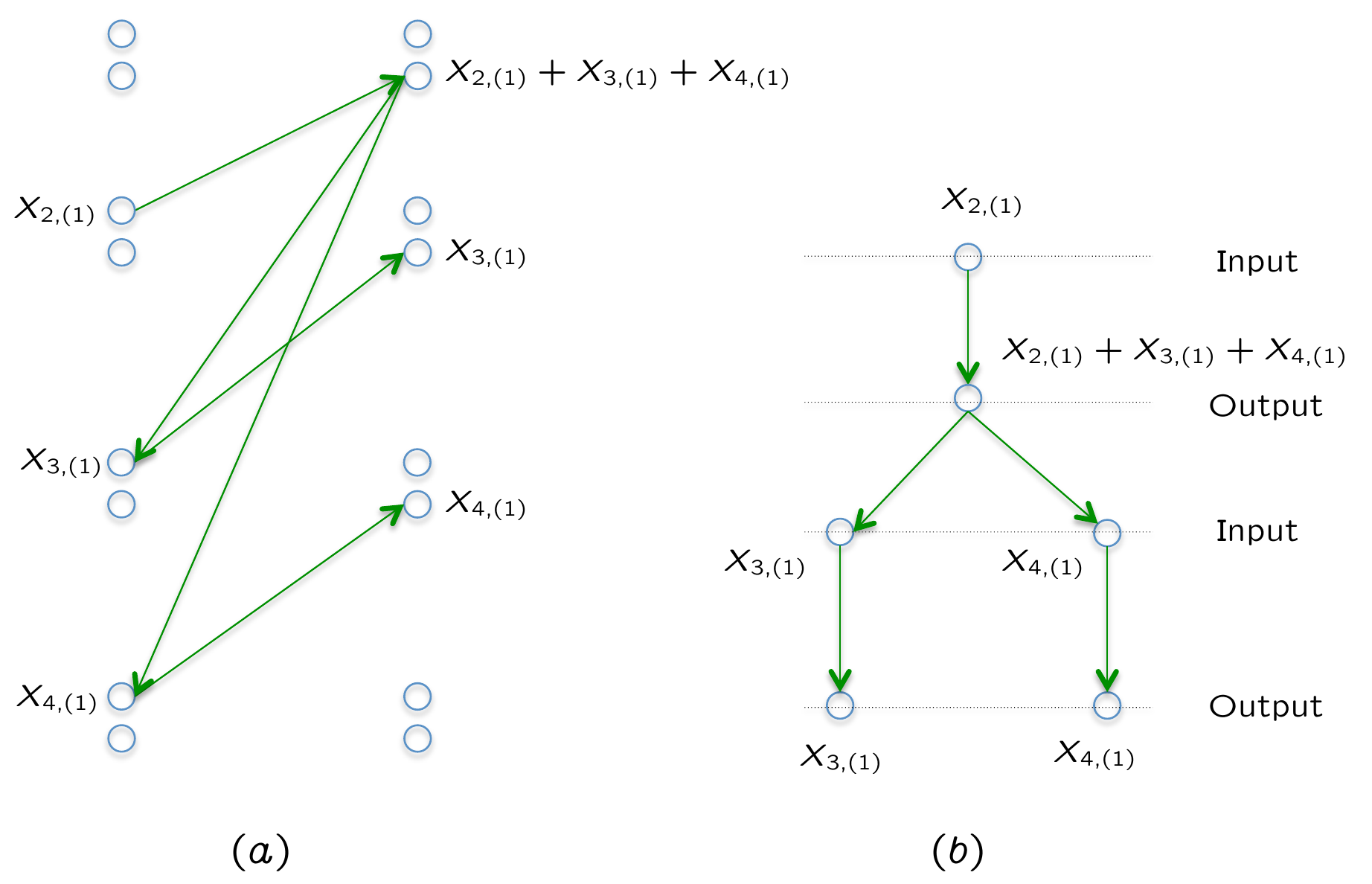}
\caption{\small $(a)$ The acyclic bipartite graph of a sub-channel that satisfies the TIN optimality condition and the tree created to invert the input bit levels. Note that the graph is undirected, direction sign is added to highlight the order of how the tree is created. $(b)$ A more tree-centric view. Note that as the graph is bipartite, the levels alternate between input and output. As participating edge is used to go from input to output, there is only one edge from an input node to an output node. The participating output level is the modulo sum of  all its connected input nodes. The leaves are output bits and are only connected to one input node from above. As such, an iterative inverting from bottom to top is feasible.}
\label{fig:inv}
\end{figure}

To illustrate the inverting process, an example would be most useful. Consider a sub-channel of a 4 user ADT deterministic interference network, whose acyclic bipartite graph is shown in Figure \ref{fig:inv}. The sub-channel is TIN optimal. Consider the optimal cyclic partition ${\Pi}^* = \{\{1,2,3,4\}\}$ with participating edges $\{e_{12}, e_{23}, e_{34}, e_{41}\}$. Let us show that it is invertible. Start from input bit $X_{2,(1)}$ and create the tree as shown in Figure \ref{fig:inv}. Inverting from the leaves would recover all input levels.


\begin{figure}[h]
\center
\includegraphics[width=5 in]{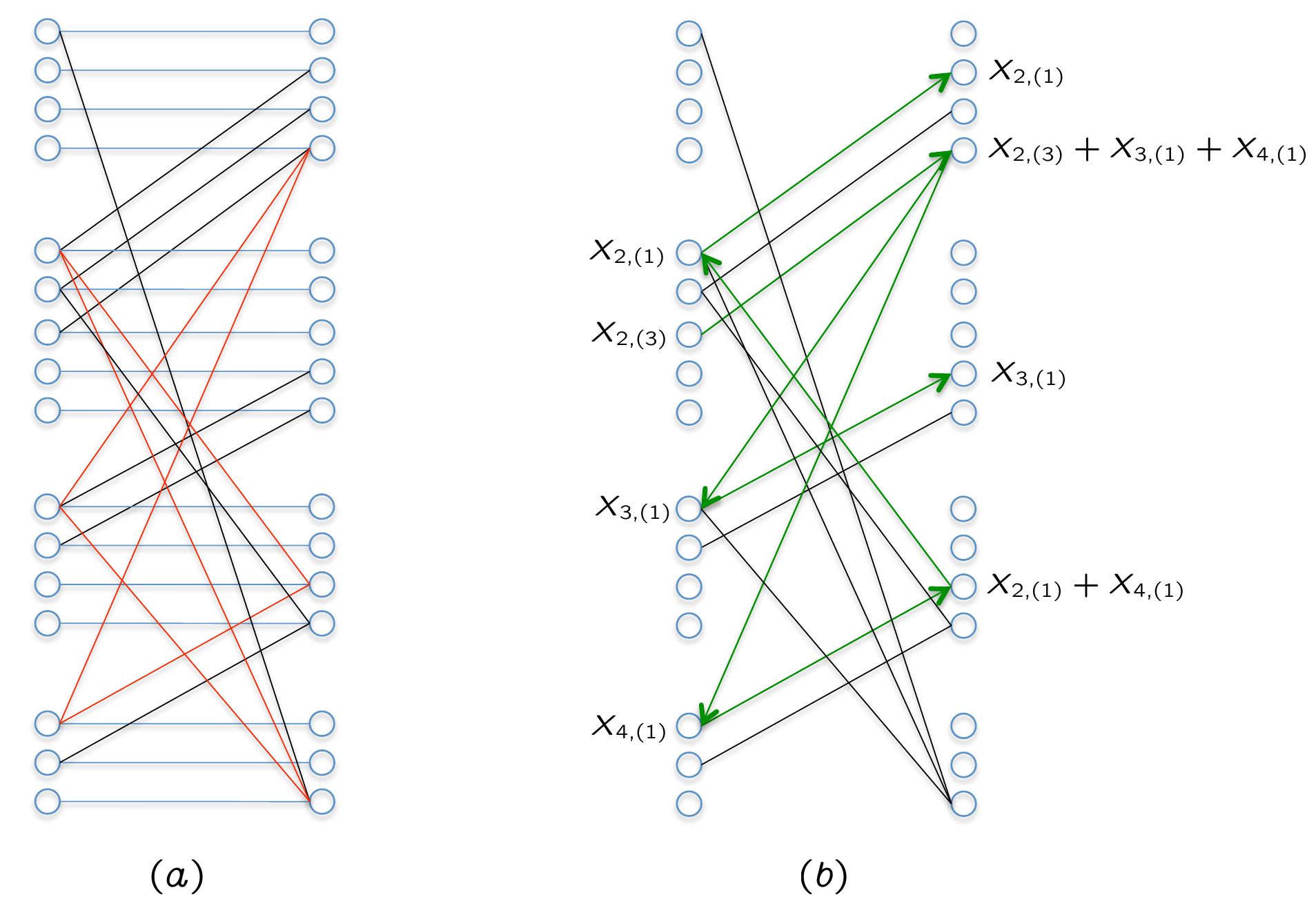}
\caption{\small $(a)$ A sub-channel that satisfies the TIN optimality condition and dominant interference condition (\ref{eq:dom}) for the dominant cyclic partition $\Pi^* = \{\{1,2,3,4\}\}$. A cycle is highlighted in red. $(b)$ The \emph{cyclic} bipartite graph and the tree created to invert $X_{2,(3)}$. Note that the graph is undirected, direction sign is added to highlight the order of how the tree is created.}
\label{fig:dom}
\end{figure}

We mention that although Theorem \ref{lemma:inv} establishes that the acyclic condition is sufficient for a sub-channel to be invertible, it is not necessary. Such examples are not uncommon, e.g., one appears in Figure \ref{fig:dom} in this paper.

Next we consider another interesting subclass of the general $K$ user ADT deterministic interference network, i.e., the cyclic interference networks where each sub-channel contains only one cycle (different sub-channels may have different cycles). As the bi-partite graph is trivially acyclic, invertibility holds. Combined with Theorem \ref{Kdetthm}, we settle the optimality of separate TIN for cyclic interference networks. The result is stated in the following corollary.

\begin{corollary}\label{col:cyc}
For a $K$ user parallel ADT deterministic interference network where each sub-channel is individually TIN optimal, if each sub-channel is also a cyclic interference network, then the sum-capacity of the $K$ user parallel ADT deterministic interference network is achieved by a separate TIN solution over each sub-channel.
\end{corollary}

{\it Remark:} Note that a \emph{cyclic} interference network has an \emph{acyclic} bi-partite graph as defined in Theorem \ref{lemma:inv}. This is because in a cyclic network each receiver receives interference from only one transmitter, so that each output level can only be connected to one input level in the bi-partite graph.

\paragraph{Networks with Dominant Partitions}
Our study of invertibility can be naturally extended to the following situation. For sub-channel $m$, consider an optimal cyclic partition ${\Pi}^{[m]*}$. If the interference caused by each Transmitter $k\in[K]$ to its cyclic predecessor $\Pi^{[m]*}(k)$ is  strictly the strongest, i.e., 
$n_{\Pi^{[m]*}(k)k}^{[m]} > n_{jk}^{[m]}, \forall j \notin\{k, \Pi^{[m]*}(k)\}$, we say that ${\Pi}^{[m]*}$ is a \emph{dominant} cyclic partition and sub-channel $m$ satisfies the \emph{dominant} interference  condition. The following theorem considers the networks where each sub-channel satisfies the dominant interference condition.

\begin{theorem}\label{dom}
For a $K$ user parallel ADT deterministic interference network where the TIN optimality condition is satisfied in each sub-channel, if each sub-channel also satisfies
\begin{eqnarray}
n_{\Pi^{[m]*}(k)k}^{[m]} > n_{jk}^{[m]},&& \forall  j,k\in[K], j \notin\{k, \Pi^{[m]*}(k)\}, m\in[M] \label{eq:dom}
\end{eqnarray}
then the sum-capacity of the  $K$ user parallel ADT deterministic interference network is achieved by a separate TIN solution over each sub-channel.
\end{theorem}

{\it Proof:} We only need to prove that when each sub-channel satisfies the dominant interference condition (\ref{eq:dom}), invertibility is implied. Although in this case, the bipartite graph may contain cycles, we are still able to construct trees in a way that no cycle would be encountered, such that inverting from the output bit leaves can recover all input levels. Similar to the construction given in Theorem \ref{lemma:inv}, for any input bit level, we leave it through a \emph{participating edge} and for any output bit level, we leave it through a \emph{non-participating} edge. When this rule is used in transversing the graph, no cycle can be created. To see this we assume the opposite. If a cycle exists when we build the tree, then each input bit node is connected to a participating edge for leaving and a non-participating edge for coming back. As this is a cycle, the net scaling factor encountered must be 0, which means the sum of the strengths of all leaving edges must equal that of all coming edges. This is a contradiction as from the dominant cyclic partition condition, for each input bit node, the strength of the leaving edge is strictly larger than that of the coming edge. So we are guaranteed to end up with a desired tree. Repeating this process would complete the proof.
\hfill\QED

We illustrate the process with an example. Consider a sub-channel of a 4 user ADT deterministic interference network, shown in Figure \ref{fig:dom}. The sub-channel is TIN optimal, as for each user, signal levels that cause interference do not suffer interference, and those that suffer interference cause no interference. Consider the optimal cyclic partition ${\Pi}^* = \{\{1,2,3,4\}\}$ with participating edges $\{e_{12}, e_{23}, e_{34}, e_{41}\}$. It is easy to verify that the participating link from each transmitter is the strongest. For example, for Transmitter 2, $n_{12} =  3 > \max(n_{32}, n_{42}) = \max(2,1) = 2$. Thus the sub-channel also satisfies the dominant interference condition. Then we prove it is invertible. Toward this end,  consider the input bit $X_{2,(3)}$. Choose the participating edge to connect to the cyclic predecessor Receiver 2. As Receiver 2 is not an end yet, we will pass through all of its non-participating edges to come to input nodes (see Figure \ref{fig:dom}). After arriving at Transmitters 3 and 4, again, follow the participating edges to cyclic predecessor Receiver 2 and 3, respectively. Receiver 2 is the end and from Receiver 3, we go to Transmitter 2 along the non-participating edge. Finally, pass through the participating edge to Receiver 1 and the end comes. It is easy to see we can invert sequentially from the output end nodes all the way to recover the desired input bit $X_{2,(3)}$ and the input bits along. All the other input bits can be recovered following similar procedures.

\subsubsection{Gaussian Setting}
We now proceed to the Gaussian setting. Starting with the 3 user case, we provide an intuitive discussion on why invertibility holds here almost surely.

\paragraph{3 users}



If the optimal cyclic partition $\Pi^*$ has two cycles, we assume $\Pi^* =\{ \{1\},\{2,3\} \}$, without loss of generality. Then $X_{1,u} = \phi, Y_{2,u} =   h_{23} X_{3,u} + n_2 , Y_{3,u} =   h_{32} X_{2,u}  +n_3 $. The participating inputs are trivially invertible from the outputs within bounded variance noise distortion here, simply by normalizing by the channel realization.

If $\Pi^*$ is a single cycle with all 3 users, we assume $\Pi^* = \{\{1,2,3\}\}$. Then $X_{1,u} = \mbox{sign}(X_1) \times 0.X_{1,(1)}\ldots X_{1,(n_{31})}, X_{2,u} = \mbox{sign}(X_2) \times 0.X_{2,(1)}\ldots X_{2,(n_{12})}, X_{3,u} = \mbox{sign}(X_3) \times 0.X_{3,(1)}\ldots X_{3,(n_{23})}$. We define $\Delta \triangleq   n_{12} + n_{23} + n_{31} - n_{21} - n_{32} - n_{13} $, which is larger than 0 almost surely for appropriately large $P$. Instead of finding a single bit as in the ADT deterministic model, we consider a chunk with $\Delta$ bits, e.g., $X_{2,[1]} = [ X_{2,(1)} \ldots X_{2,(\min(\Delta, n_{12}))} ]$. Operating in units of $\Delta$ bits, the invertibility process parallels the ADT deterministic model. The effect of additive noise terms becomes vanishingly small at the higher signal levels (thus limited to only an $o(\log P)$ impact, see \cite{Cadambe_Jafar_fullyconnected} for this argument), the carry overs across chunks are vanishingly small relative to the size of the chunks, and their number also does not scale with $P$ because the number of chunks remains constant. Thus, the Gaussian setting parallels the deterministic setting within $o(\log P)$.  Note that as the  condition for non-invertibility in the  ADT deterministic model is approached, i.e., as $\alpha_{12} + \alpha_{23} + \alpha_{31} - \alpha_{21} - \alpha_{32} - \alpha_{13}$ approaches zero, the size of the chunks becomes smaller, and the overhead of carry over bits increases proportionately. However, except when it is exactly zero (the setting with infinite overhead), the overhead does not scale with $P$, thus the GDoF, almost surely, continue to mimic the deterministic setting.

\paragraph{Networks with Dominant Partitions}

\begin{theorem}\label{gdom}
For a $K$ user parallel Gaussian interference network where the TIN optimality condition is satisfied in each sub-channel, if each sub-channel also satisfies
\begin{eqnarray}
\alpha_{\Pi^{[m]*}(k)k}^{[m]} > \alpha_{jk}^{[m]},&& \forall  j,k\in[K], j \notin\{k, \Pi^{[m]*}(k)\}, m\in[M] \label{eq:gdom}
\end{eqnarray}
then the sum-GDoF value of the $K$ user parallel Gaussian interference network is achieved by a separate TIN solution over each sub-channel.
\end{theorem}

{\it Proof:}  Instead of an appeal to the ADT deterministic model, which could still be made, it is worthwhile in this section to consider a more direct proof. So let us see why the invertibility property is almost surely true, i.e., $H({\bf X}_u^{[m]} | {\bf Y}_u^{[m]}) = o(\log P).$
We focus on one sub-channel, and the sub-channel index is omitted. We have
\begin{eqnarray}
- I({\bf X}_u ; {\bf Y}_u) 
&=& H({\bf X}_u| {\bf Y}_u) - H({\bf X}_u) \\
&=& \underbrace{ h({\bf Y}_u | {\bf X}_u) }_{o(\log P)} - h({\bf Y}_u)\\
\Longleftrightarrow H({\bf X}_u | {\bf Y}_u) &=& H({\bf X}_u)  - h({\bf Y}_u) + o(\log P)
\end{eqnarray}
Thus, for invertibility, it suffices to prove  $H({\bf X}_u)  - h({\bf Y}_u) = o(\log P).$

We prove that when (\ref{eq:gdom}) holds for sub-channel $m$, the invertibility property is implied.
Towards this end, we define $V_{i,u} = h_{\Pi^{[m]*}(i)i} X_{i,u} + Z_{\Pi^{[m]*}(i)}$, ${\bf V}_{u} = (V_{1,u}, \ldots, V_{K,u})$ and prove
\begin{eqnarray}
H({\bf X}_u)  - h({\bf V}_u) &=& o(\log P) \label{inve1} \\
h({\bf V}_u)  - h({\bf Y}_u) &=& o(\log P) \label{inve2} .
\end{eqnarray}
Let us prove them one by one. First, consider (\ref{inve1}). It can be proved by noticing that $|h_{ \Pi^{[m]*}(i)  i}| = \sqrt{P^{\alpha_{\Pi^{[m]*}(i)i} }}$ such that in ${\bf V}_u$, all bits in ${\bf X}_u$ are received above the noise floor. The derivations are similar to those in \cite{Bresler_Tse,Cadambe_Jafar_fullyconnected},  thus we omit it.

Next, we prove (\ref{inve2}). Let us rewrite ${\bf V}_u$ and ${\bf Y}_u$ in the matrix form
\begin{eqnarray}
{\bf V}_u = {\bf G} {\bf {X}}_u + {\bf {\bar{Z}}}, {\bf Y}_u = {\bf F} {\bf X}_u + {\bf Z}
\end{eqnarray}
where
\begin{eqnarray}
{\bf G} &=& \mbox{diag} ( h_{ \Pi^{[m]*}(1) 1}, h_{ \Pi^{[m]*}(2) 2} , \ldots, h_{ \Pi^{[m]*}(K) K}) \\
{\bf F} &=& [h_{ji}]_{K \times K} - \mbox{diag}(h_{11}, \ldots, h_{KK})
\end{eqnarray}
and ${\bf {\bar{Z}}} = (Z_{\Pi^{[m]*}(1) }, \ldots, Z_{\Pi^{[m]*}(K) })$ is a permutation of ${\bf {{Z}}} = (Z_1, \ldots, Z_K)$.
${\bf G}$ and ${\bf F}$ are invertible almost surely and
\begin{eqnarray}
{\bf F}{\bf G}^{-1} &=& \left[\frac{h_{ji}}{h_{\Pi^{[m]*}(i) i}}\right]_{K \times K} - \mbox{diag}\left(\frac{h_{11}}{h_{\Pi^{[m]*}(1)1}}, \ldots, \frac{h_{KK}}{h_{\Pi^{[m]*}(K) K}}\right)
\end{eqnarray}

Define $\sigma$ as the smallest singular value of ${\bf F} {\bf G}^{-1}$, and introduce $\beta \triangleq \min(\sigma, 1)$. Let us also define ${\bf Z'} \sim \mathcal{N} (0, {\bf F}{\bf G}^{-1} ({\bf F} {\bf G}^{-1})^{T} - \beta {\bf I})$ and ${\bf Z'}$ is independent of ${\bf Z}$. The positive semidefinite property of the covariance
matrix is easily established from the definition of $\beta$. We now have
\begin{eqnarray}
&& h({\bf V}_u)  - h({\bf Y}_u) \notag\\
&=& h({\bf G} {\bf {X}}_u + {\bf {\bar{Z}}} ) - h(  {\bf F} {\bf {X}}_u + {\bf {{Z}}}) \\
&\leq& h({\bf G} {\bf {X}}_u + {\bf {\bar{Z}}}) - h(  {\bf F} {\bf {X}}_u + \beta{\bf {{Z}}})  \label{z} \\
&=& h({\bf G} {\bf {X}}_u + {\bf {\bar{Z}}} ) -  I( {\bf F} {\bf {X}}_u + \beta{\bf {{Z}}} ; {\bf F} {\bf {X}}_u) - h(\beta {\bf Z})\\
&\leq& h({\bf G} {\bf {X}}_u + {\bf {\bar{Z}}} ) -  I( {\bf F} {\bf {X}}_u + \beta {\bf {{Z}}} + {\bf Z}'; {\bf F} {\bf {X}}_u) - h(\beta {\bf Z}) \label{data}\\
&=& h({\bf G} {\bf {X}}_u + {\bf {\bar{Z}}} ) -  \underbrace{h( {\bf F} {\bf {X}}_u + \beta {\bf {{Z}}} + {\bf Z'})}_{= h\left( {\bf  F} {\bf  G}^{-1} ({\bf G} {\bf {X}}_u + {\bf {\bar{Z}}}) \right)} + h(\beta {\bf {{Z}}} + {\bf Z'}) - h(\beta {\bf Z}) \\
&=& h({\bf G} {\bf {X}}_u + {\bf {\bar{Z}}} ) - h({\bf G} {\bf {X}}_u + {\bf {\bar{Z}}} ) - \log \left| {\bf  F} {\bf  G}^{-1} \right|\\
&& +~ \frac{1}{2} \log(2 \pi e)^K \left| {\bf  F} {\bf  G}^{-1} ({\bf  F} {\bf  G}^{-1})^T  - \beta^2 {\bf I} + \beta^2 {\bf I} \right|  - \frac{1}{2} \log(2 \pi e)^K |\beta^2 {\bf I} |\\
&=& - \log \left| {\bf  F} ({\bf  G})^{-1} \right| + \frac{1}{2} \log|{{\bf  F} \bf  G}^{-1}| |({\bf  F} {\bf  G}^{-1})^T |
- \frac{1}{2} \log (\beta^2) \\
&=&  -\frac{1}{2} \log (\beta^2) \label{st}
\end{eqnarray}
where (\ref{z}) follows from the fact that $\beta\leq1$. In (\ref{data}), we use the data processing inequality as $ {\bf F} {\bf X}_u \rightarrow {\bf F} {\bf X}_u + \beta{\bf Z} \rightarrow{\bf F} {\bf X}_u + \beta {\bf Z} + {\bf Z'}$ forms a Markov chain. 

It only remains to show that $\beta$ is $o(\log P)$. As $\beta = \min(\sigma,1)$, it suffices to show $\sigma = o(\log P)$. By definition, $\sigma = \min_{{\bf x}} || {\bf F} {\bf G}^{-1} {\bf x}||$, where ${\bf x} \in {\mathbb R}^{K \times 1}$ is a unit vector. 
Let us prove the claim by contradiction. Choose a small positive $\epsilon$ such that $\epsilon^2 < \frac{1}{2K^3}$. Suppose $\sigma$ decays too fast with respect to $P$, then choose $P$ sufficiently large such that
\begin{eqnarray}
\sigma = \min_{||{\bf x}||=1}  ||{\bf F} {\bf G}^{-1} {\bf x}|| &\leq& \epsilon \label{yy}\\
\frac{|h_{ji}|}{|h_{\Pi^{[m]*}(i)  i}|} = \sqrt{P^{  \alpha_{ji} - \alpha_{\Pi^{[m]*}(i)  i} } }&\leq& \epsilon, \forall j \notin \{ i, \Pi^{[m]*}(i) \}. \label{small}
\end{eqnarray}
Suppose the minimizing unit vector that corresponds to $\sigma$ is ${\bf x^*} = [x_1, \ldots, x_K]^T$. Then the $j$-th entry of the $K \times 1$ vector ${\bf F} {\bf G}^{-1} {\bf x^*}$ (denoted as $y_j$) is 
\begin{align}
y_j = \sum_{i=1, i \neq j}^K \frac{h_{ji}}{h_{\Pi^{[m]*}(i) i}} x_i = \sum_{i=1, i \neq j, \Pi^{[m]*}(i)  \neq j}^K \frac{h_{ji}}{h_{\Pi^{[m]*}(i) i}} x_i + x_{i_o}
\end{align}
where $\Pi^{[m]*}(i_o) = j$ and its absolute value
\begin{align}
|y_j| &\geq  |x_{i_o}| -  \sum_{i=1, i \neq j, \Pi^{[m]*}(i)  \neq j}^K \left| \frac{h_{ji}}{h_{\Pi^{[m]*}(i) i}} x_i \right| \\
& \geq  |x_{i_o}| - (K-2) \epsilon \label{ee}
\end{align}
where (\ref{ee}) follows from (\ref{small}) and $|x_i| \leq 1$ as ${\bf x}^*$ is a unit vector. Also,
\begin{align}
1 = \sum_{i_o = 1}^K |x_{i_o}|^2 &\leq \sum_{j=1}^K \Big( |y_j| + (K-2) \epsilon \Big)^2 \label{yj} \\
&\leq \sum_{j=1}^K 2 \Big( |y_j|^2 + (K-2)^2 \epsilon^2 \Big) \\
&\leq 2 \epsilon^2 + 2K(K-2)^2 \epsilon^2 \leq 2K^3 \epsilon^2 \label{ss} 
\end{align}
where we use (\ref{ee}) to get (\ref{yj}) and (\ref{yy}) is used in (\ref{ss}) such that $\sum_{j=1}^K |y_j|^2 \leq \epsilon^2$. We get the desired contradiction as $\epsilon^2 < \frac{1}{2K^3}$ by assumption.\hfill\QED

The above proof relies heavily on the fact that $\sigma = o(\log P)$, which is only true when the dominant interference condition is satisfied. In general, we can not use this direct matrix inversion method to prove the invertibility for the Gaussian case. Proofs along the lines of the ADT deterministic model seem more generally applicable.


\subsection{GDoF Region for Parallel Networks}\label{sec:SFregion}
The GDoF region of a TIN optimal $K$ user interference network, as stated in Theorem \ref{theorem:Geng}, is comprised only of   sum-GDoF  bounds for all subsets of users. For parallel TIN optimal interference networks, our results characterize the tight sum-GDoF bounds of any subset of users. So it is natural to wonder if the set of  all tight sum-GDoF bounds for all subsets of users  characterizes the entire GDoF region, and therefore settles the optimality of TIN for the entire GDoF \emph{region} in the \emph{parallel} setting. In this section, we show through a counter-example that this is not the case. The following theorem states the result.

\begin{theorem}\label{theorem:gap}
For the parallel $K>2$ user Gaussian interference network with $m>1$ sub-channels, each of which is individually TIN optimal and invertible,  the region described by the tightest sum-GDoF bounds of all subsets of users, is in general \emph{not} the same as the region achievable by separate TIN over each sub-channel.
\end{theorem}

\noindent{\it Remark: Note that if either $K=2$ or $m=1$, then the two regions are the same. When $K>2$ and $m>1$, even though the regions are not the same, the sum-GDoF values are indeed the same, as we have shown in Theorem \ref{thm}. Theorem \ref{theorem:gap} also applies to the ADT deterministic model. This is readily seen because the counter-example presented below extends to the deterministic setting by choosing integer values $n_{ij}^{[m]}=10\alpha_{ij}^{[m]}$, $\forall i,j\in\{1,2,3\}, m \in \{1,2\}$, $\epsilon=1$.}


\begin{figure}[h]
\center
\includegraphics[width= 3 in]{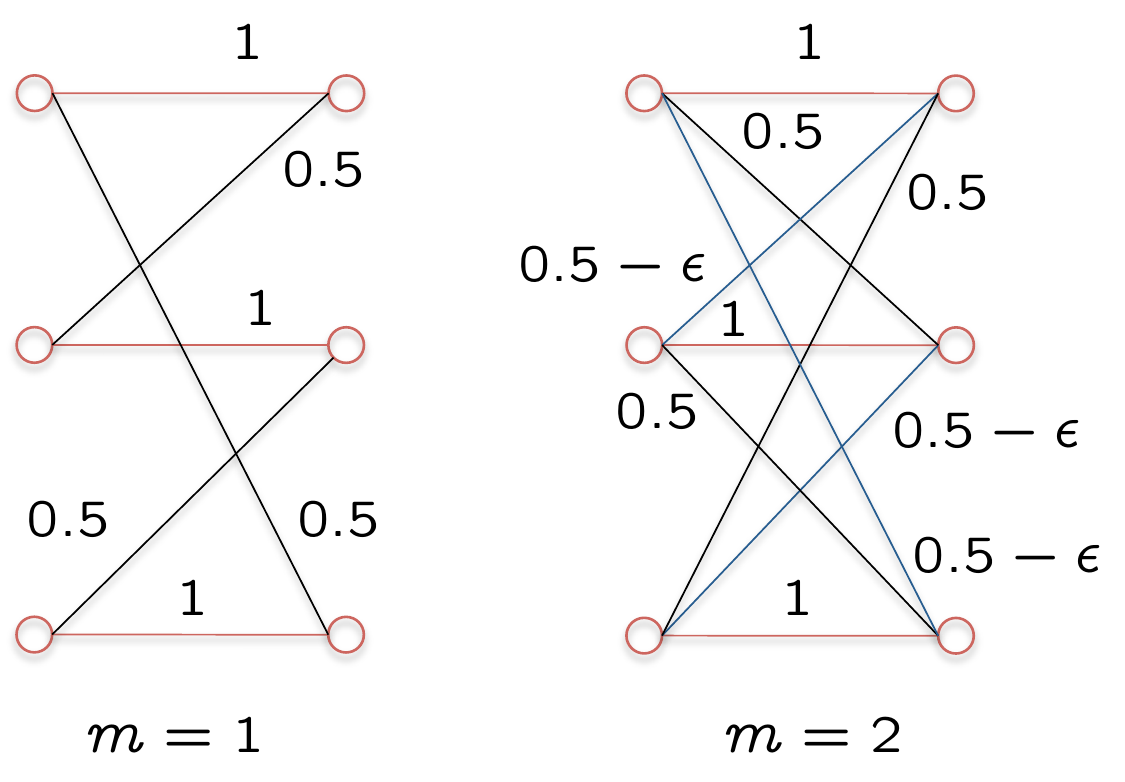}
\caption{\small A $K=3$ user Gaussian interference network with 2 sub-channels. The channel strength level is indicated for each link. Each sub-channel satisfies the TIN optimal condition and dominant interference condition.}
\label{reg}
\end{figure}
\proof
Consider a $K = 3$ user Gaussian interference network with $M = 2$ sub-channels, as shown in Figure \ref{reg}. 
It is easily seen that both sub-channels satisfy the TIN optimality condition and the dominant interference condition, for all subsets of users. Therefore, Theorem \ref{gdom} establishes that the sum-GDoF value of  all subsets of users in this parallel Gaussian interference network  is achieved by separate TIN over each sub-channel. Incidentally, the sum-GDoF value for all $3$ users is 3, achieved by the GDoF tuple $(d_1,d_2,d_3) = (1,1,1)$ where every user gets 0.5 GDoF over each sub-channel by transmitting at full power and each receiver treats interference as noise.

We now view each TIN optimal sub-channel by itself. The GDoF region of the first sub-channel by itself is the set of tuples $(d_1^{[1]}, d_2^{[1]}, d_3^{[1]}) \in \mathbb{R}_{+}^3$ defined by the following constraints. 

\begin{eqnarray}
 d_i^{[1]} &\leq& 1, ~~\forall i \in \{1,2,3\}\\
d_i^{[1]} + d_j^{[1]} &\leq& 1.5, ~~\forall i, j \in \{1,2,3\}, i \neq j \label{zs} \\
d_1^{[1]}  + d_2^{[1]}  + d_3^{[1]}  &\leq& 1.5
\end{eqnarray}

Similarly, the individual GDoF region for the second sub-channel is 
\begin{eqnarray}
 d_i^{[2]} &\leq& 1, ~~\forall i \in \{1,2,3\}\\
d_i^{[2]} + d_j^{[2]} &\leq& 1 + \epsilon, ~~\forall i, j \in \{1,2,3\}, i \neq j \label{f1} \\
d_1^{[2]}  + d_2^{[2]}  + d_3^{[2]}  &\leq& 1.5
\end{eqnarray}

Considering all sub-channels together, the sum-GDoF bounds for the parallel interference network (each of which is tight by itself, as proved in Theorem \ref{thm}) are the following.
\begin{eqnarray}
{d}_i &\leq& 1+ 1 = 2, ~~\forall i \in \{1,2,3\}\label{eq:c1}\\
{d}_i + {d}_j &\leq& 1.5 + 1 + \epsilon = 2.5 + \epsilon, ~~\forall i, j \in \{1,2,3\}, i \neq j\\
{d}_1  + {d}_2  + {d}_3 &\leq&  1.5 + 1.5 = 3\label{eq:c3}
\end{eqnarray}

Now, consider the GDoF tuple  $({d}_1, {d}_2, {d}_3) = (2, 0.5, 0.5)$ which is inside the region described by (\ref{eq:c1})-(\ref{eq:c3}). We prove this tuple is not achievable by separate TIN. In other words, we show that there does not exist a valid $(d_1^{[1]}, d_2^{[1]}, d_3^{[1]})$ and a valid $(d_1^{[2]}, d_2^{[2]}, d_3^{[2]})$, such that $(d_1^{[1]}+d_1^{[2]}, d_2^{[1]}+d_2^{[2]}, d_3^{[1]}+d_3^{[2]}) = (2, 0.5, 0.5)$. This is shown as follows.

In order to have $d_1^{[1]}+d_1^{[2]} = 2$,  we must have $d_1^{[1]} = d_1^{[2]}  = 1$. Given $d_1^{[2]} = 1$, from (\ref{f1}), we must have  $d_2^{[2]}\leq\epsilon$ and $d_3^{[2]} \leq \epsilon$. Since $d_2^{[2]}\leq\epsilon$, then, in order to have $d_2^{[1]}+d_2^{[2]}=0.5$, we must have 
$d_2^{[1]} \geq 0.5 -  \epsilon$. 
Since $d^{[1]}_1=1$, $d_2^{[1]}\geq 0.5-\epsilon$ and $d_1^{[1]}+d^{[1]}_2+d^{[1]}_3\leq 1.5$, we must have $d^{[1]}_3\leq \epsilon$. Now, since $d^{[1]}_3\leq\epsilon$ and $d^{[2]}_3\leq\epsilon$, we must have $d^{[1]}_3+d^{[2]}_3\leq 2\epsilon$. And since $\epsilon>0$ can be arbitrarily small, it contradicts the requirement that $d^{[1]}_3+d^{[2]}_3=0.5$, thus completing the proof by counter-example.\hfill\QED

To summarize, for  parallel interference networks (deterministic and Gaussian), where each sub-channel is individually TIN optimal and invertible, either the separate TIN achievable region is not tight or we need more than sum-rate bounds. In light of this observation, the optimality of separate TIN for sum-GDoF is especially remarkable.

\section{Proofs}
\subsection{Redundancy of non-negativity constraints in $LP_1$}\label{sec:sign}
Before we prove the redundancy of non-negativity constraints in $LP_1$, let us first highlight the non-trivial nature of the problem. Consider the following $LP$, which seems similar to $LP_1$.
$$\max{R_1+R_2+R_3~\mbox{ such that }~ R_1+R_2\leq 10, R_1+R_3\leq 10, R_2+R_3\leq 30, (R_1, R_2, R_3)\in\mathbb{R}^3_+}$$
It is easy to see that the max value is $20$ achieved with $(R_1,R_2,R_3)=(0,10,10)$. However, if we ignore the non-negativity constraint $(R_1, R_2, R_3)\in\mathbb{R}^3_+$, then we can achieve a sum value of 25 with $(R_1,R_2,R_3)=(-5,15,15)$. Thus, in this $LP$, which looks similar to $LP_1$, one \emph{cannot} ignore the non-negativity constraints. So let us see why this can be done in $LP_1$.

Returning to sum-GDoF characterization in $LP_1$, we already assumed that the TIN optimality condition (\ref{eq:cond}) is  satisfied by the network, but let us now further assume that it is satisfied with \emph{strict} inequality. We note that there is no loss of generality here, because the case with equality  immediately follows from a continuity argument. Strict inequality in the TIN optimality condition means the following is true.
\begin{equation}
\alpha_{ii}> \max_{j:j\neq i}\{\alpha_{ji}\}+\max_{k:k\neq i}\{\alpha_{ik}\},~~~\forall i,j,k\in[K] \label{eq:positiverate}
\end{equation}
We need the following lemmas.
\begin{lemma}\label{lemma:positiverate}
Given that (\ref{eq:positiverate}) is satisfied, the sum-GDoF must be achieved by a GDoF tuple $(d_1, d_2, \cdots, d_K)$ with  $d_k>0, \forall k\in[K]$.
\end{lemma}
\proof Suppose  that the sum-GDoF are achieved with a GDoF tuple where $d_i=0$. Replacing $n_{ij}$ with $\alpha_{ij}$ in Fig. \ref{fig:cond}, it is evident that  user $i$ has $\alpha_{ii}- \max_{j:j\neq i}\{\alpha_{ji}\}-\max_{k:k\neq i}\{\alpha_{ik}\}$ signal levels that neither cause interference, nor suffer interference. Thus, user $i$ can be assigned $d_i = \alpha_{ii}- \max_{j:j\neq i}\{\alpha_{ji}\}-\max_{k:k\neq i}\{\alpha_{ik}\}>0$ GDoF without hurting any other user, thus improving the sum-GDoF value. Since the sum-GDoF value cannot be improved, we have a contradiction that completes the proof.\hfill$\Box$

\begin{lemma}\label{lemma:drop}
Consider a region
\begin{eqnarray}
\mathcal{D}&=&\mathbb{R}^K_+\cap\mathcal{D}_u
\end{eqnarray}
where $\mathcal{D}_u\subset\mathbb{R}^K$ is closed and convex. If $\max_{(d_1, d_2, \cdots, d_K)\in \mathcal{D}}\sum_{k=1}^Kd_k\define S<\infty$ is achieved by a tuple $(d_1, d_2,\cdots, d_K)$ with  $d_k>0, \forall k\in[K]$, then 
\begin{eqnarray}
\max_{(d_1, d_2, \cdots, d_K)\in \mathcal{D}}\sum_{k=1}^Kd_k&=&
\max_{(d_1, d_2, \cdots, d_K)\in \mathcal{D}_u}\sum_{k=1}^Kd_k
\end{eqnarray}
\end{lemma}
\proof To set up a proof by contradiction, suppose, on the contrary, that while the $\max$ sum value in $\mathcal{D}$ is $S$, which is achieved by the tuple ${\bf d}\in\mathcal{D}$ with $d_k>0, \forall k\in[K]$, there exists a tuple ${\bf d}_u\in\mathcal{D}_u$ that achieves the sum value $S_u$ such that $S<S_u<\infty$. Define ${\bf v}={\bf d}_u - {\bf d}$ and $S_v=\sum_{k=1}^{K}v_k$. Clearly, $S_v=S_u-S>0$. Consider the tuple ${\bf d}_\epsilon={\bf d}+\epsilon {\bf v}$, with $\epsilon\in[0,1]$, chosen  such that ${\bf d}_\epsilon\in\mathbb{R}^K_+$. This is possible because all elements of ${\bf d}$ are strictly positive. Since $\mathcal{D}_u$ is convex, and we have both ${\bf d}\in\mathcal{D}_u$ and ${\bf d}_u\in\mathcal{D}_u$, therefore we must have  a convex combination of the two, ${\bf d}_\epsilon\in\mathcal{D}_u$. Since we also have ${\bf d}_\epsilon\in\mathbb{R}^K_+$, it follows that ${\bf d}_\epsilon\in\mathcal{D}$. But this is a contradiction, because the sum-value achieved by ${\bf d}_\epsilon$ is $S_\epsilon=S+\epsilon S_v>S$, when $S$ was assumed to be the max value in $\mathcal{D}$. \hfill$\Box$


By choosing $\mathcal{D}$ as the constraint space for $LP_1$, and $\mathcal{D}_u$ as the same region without the non-negativity constraint on the $d_i$, Lemma \ref{lemma:positiverate} and Lemma \ref{lemma:drop} imply that if (\ref{eq:positiverate}) is satisfied, then there is no loss of generality in dropping the non-negativity constraints in $LP_1$. 

Finally in the case where the TIN optimality condition is satisfied possibly with equalities, a simple continuity argument can be applied as follows. Let us increase all $\alpha_{ii}$ by a small positive amount $\epsilon$. The resulting network is still TIN optimal, but now it satisfies the TIN optimality condition with a strict inequality. Since each of the bounds is perturbed by at most $K\epsilon$, the sum-GDoF for the new network cannot exceed that of the original by more than $K\epsilon$. Note that for the new network, because of Lemma \ref{lemma:positiverate} and Lemma \ref{lemma:drop} one can drop the non-negativity constraints with no loss of generality. Thus, in the limit $\epsilon\rightarrow 0+$, the sum-GDoF of the old network and the new network converge to the same value, as do the two linear programs, with and without the non-negativity constraints. Thus, even when the users satisfy only the TIN optimality condition (\ref{eq:cond}) there is no loss of generality in dropping the non-negativity constraints. 

\subsection{Outer Bound Proof of Example 1}
\begin{figure}[h]
\center
\includegraphics[width=6in]{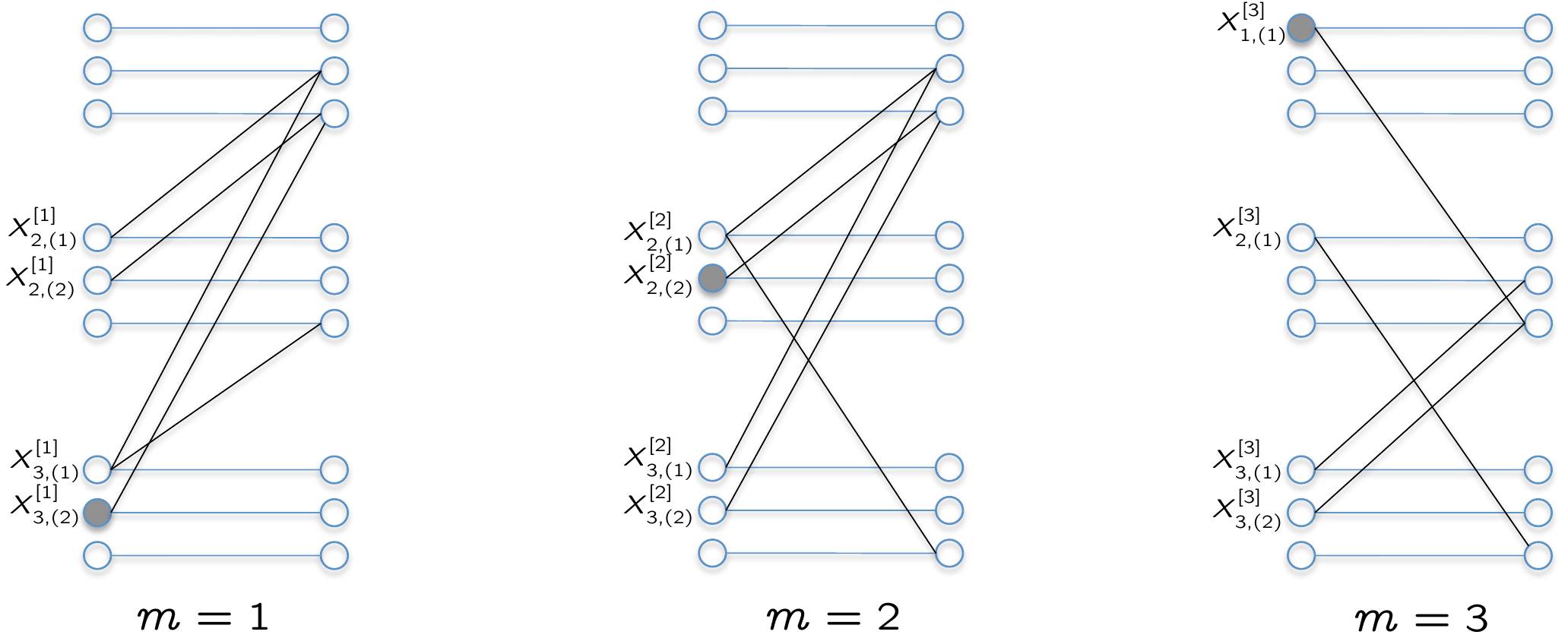}
\caption{\small The same 3 user parallel ADT deterministic interference network network as Example 1. All the interfering input bits are labeled. Those that do \emph{not} belong to $X_{i,u}^{[m]}$ are made solid.}
\label{pex1}
\end{figure}

Before going to the outer bound proof for Theorem \ref{Kdetthm}, we provide a proof specifically for Example 1 first, in order to illustrate the main insights in a simpler setting. For clarity of exposition, we redraw the network in Figure \ref{pex1}. We want to prove the sum-capacity of this 3 user parallel ADT deterministic interference network is bounded above by $18$.

For Receiver 1, from Fano's inequality, we have
\begin{eqnarray}
n(R_1 - \epsilon) &\leq& I(W_1; Y_1^{[1]^n}, Y_1^{[2]^n}, Y_1^{[3]^n}) \\
&=& H(Y_1^{[1]^n}, Y_1^{[2]^n}, Y_1^{[3]^n}) - H(Y_1^{[1]^n}, Y_1^{[2]^n}, Y_1^{[3]^n} | W_1) \\
&\leq& 9n - H(X_{2,(1)}^{[1]^n} \oplus X_{3,(1)}^{[1]^n}, X_{2,(2)}^{[1]^n} \oplus X_{3,(2)}^{[1]^n}, X_{2,(1)}^{[2]^n} \oplus X_{3,(1)}^{[2]^n}, X_{2,(2)}^{[2]^n} \oplus X_{3,(2)}^{[2]^n}) \label{e1}
\end{eqnarray}
where (\ref{e1}) follows from the fact that each bit can only carry at most 1 bit of information.

For Receiver 2, we provide the bits that are sent from Transmitter 2 and cause interference at undesired receivers, i.e., the bits labeled in Figure \ref{pex1}, as side information from a genie. Then we have
\begin{eqnarray}
&& n(R_2 - \epsilon) \notag \\
&\leq& I(W_2; Y_2^{[1]^n}, Y_2^{[2]^n}, Y_2^{[3]^n}, X_{2,(1)}^{[1]^n}, X_{2,(2)}^{[1]^n}, X_{2,(1)}^{[2]^n}, X_{2,(2)}^{[2]^n}, X_{2,(1)}^{[3]^n}) \\
&=& H(Y_2^{[1]^n}, Y_2^{[2]^n}, Y_2^{[3]^n}, X_{2,(1)}^{[1]^n}, X_{2,(2)}^{[1]^n}, X_{2,(1)}^{[2]^n}, X_{2,(2)}^{[2]^n},X_{2,(1)}^{[3]^n}) \notag \\
&& -~ H(Y_2^{[1]^n}, Y_2^{[2]^n}, Y_2^{[3]^n}, X_{2,(1)}^{[1]^n}, X_{2,(2)}^{[1]^n}, X_{2,(1)}^{[2]^n}, X_{2,(2)}^{[2]^n},X_{2,(1)}^{[3]^n} | W_2) \\
&=& H(X_{2,(2)}^{[2]^n}) + H(X_{2,(1)}^{[1]^n}, X_{2,(2)}^{[1]^n}, X_{2,(1)}^{[2]^n},X_{2,(1)}^{[3]^n} | X_{2,(2)}^{[2]^n}) \notag \\
&& +~ \underbrace{ H(Y_2^{[1]^n}, Y_2^{[2]^n}, Y_2^{[3]^n} | X_{2,(1)}^{[1]^n}, X_{2,(2)}^{[1]^n}, X_{2,(1)}^{[2]^n}, X_{2,(2)}^{[2]^n},X_{2,(1)}^{[3]^n})}_{\leq 4n} \notag \\
&& - H(X_{3,(1)}^{[1]^n}, X_{3,(1)}^{[3]^n}, X_{3,(2)}^{[3]^n}  \oplus X_{1,(1)}^{[3]^n} ) \\
&\leq& 5n + H(X_{2,(1)}^{[1]^n}, X_{2,(2)}^{[1]^n}, X_{2,(1)}^{[2]^n},X_{2,(1)}^{[3]^n} | X_{2,(2)}^{[2]^n})  - H(X_{3,(1)}^{[1]^n}, X_{3,(1)}^{[3]^n}, X_{3,(2)}^{[3]^n} \oplus X_{1,(1)}^{[3]^n})  \label{e2}
\end{eqnarray}
where in (\ref{e2}), the positive term is exactly ${\bf X}_{2,u}$ with conditioning on other interfering bit (solid node in Figure \ref{pex1}), and the negative term is the interference. Similarly, for Receiver 3, we have
\begin{eqnarray}
n(R_3 - \epsilon) &\leq& 4n + \underbrace{ H(X_{3,(1)}^{[1]^n}, X_{3,(1)}^{[2]^n}, X_{3,(2)}^{[2]^n},X_{3,(1)}^{[3]^n}, X_{3,(2)}^{[3]^n} | X_{3,(2)}^{[1]^n})}_{=H( {\bf X}_{3,u}^n | X_{3,(2)}^{[1]^n}) }  - \underbrace{H(X_{2,(1)}^{[2]^n}, X_{2,(1)}^{[3]^n})}_{\mbox{Interference}}.  \label{e3}
\end{eqnarray}
Adding (\ref{e1}), (\ref{e2}) and (\ref{e3}), we have
\begin{eqnarray}
n(R_1 + R_2 + R_3 - \epsilon) &\leq& 18n + H( {\bf X}_{2,u}^n | X_{2,(2)}^{[2]^n}) + H( {\bf X}_{3,u}^n | X_{3,(2)}^{[1]^n}) \notag \\
&& -~ H(X_{2,(1)}^{[1]^n} \oplus X_{3,(1)}^{[1]^n}, X_{2,(2)}^{[1]^n} \oplus X_{3,(2)}^{[1]^n}, X_{2,(1)}^{[2]^n} \oplus X_{3,(1)}^{[2]^n}, X_{2,(2)}^{[2]^n} \oplus X_{3,(2)}^{[2]^n}) \notag \\
&&-~ H(X_{3,(1)}^{[1]^n}, X_{3,(1)}^{[3]^n}, X_{3,(2)}^{[3]^n} \oplus X_{1,(1)}^{[3]^n} ) - H(X_{2,(1)}^{[2]^n}, X_{2,(1)}^{[3]^n}) \\
&\leq& 18n + H( {\bf X}_{2,u}^n, {\bf X}_{3,u}^n | X_{1,(1)}^{[3]^n}, X_{2,(2)}^{[2]^n}, X_{3,(2)}^{[1]^n}) \notag \\
&& -~ H(X_{2,(1)}^{[1]^n} \oplus X_{3,(1)}^{[1]^n}, X_{2,(2)}^{[1]^n}, X_{2,(1)}^{[2]^n} \oplus X_{3,(1)}^{[2]^n}, X_{3,(2)}^{[2]^n},\ldots \notag \\
&& X_{3,(1)}^{[1]^n}, X_{3,(1)}^{[3]^n}, X_{3,(2)}^{[3]^n}, X_{2,(1)}^{[2]^n}, X_{2,(1)}^{[3]^n} | X_{1,(1)}^{[3]^n}, X_{2,(2)}^{[2]^n}, X_{3,(2)}^{[1]^n}) \label{e4} \\
&=& 18n + H( {\bf X}_{1,u}^n, {\bf X}_{2,u}^n, {\bf X}_{3,u}^n | X_{1,(1)}^{[3]^n}, X_{2,(2)}^{[2]^n}, X_{3,(2)}^{[1]^n}) \notag \\
&& -~ H( {\bf Y}_{1,u}^n, {\bf Y}_{2,u}^n, {\bf Y}_{3,u}^n | X_{1,(1)}^{[3]^n}, X_{2,(2)}^{[2]^n}, X_{3,(2)}^{[1]^n}) \\
&=& 18n
\end{eqnarray}
where in (\ref{e4}), the second term follows from the independence of ${\bf X}_i$ and in the third term, we add conditioning on $X_{1,(1)}^{[3]}, X_{2,(2)}^{[2]}, X_{3,(2)}^{[1]}$, which cannot increase entropy. The negative term in (\ref{e4}) is now the interfering signals resulting from ${\bf X}_{i,u}$, i.e., ${\bf Y}_{i,u}$. In the last step, we use the invertibility property, already verified for Example 1. Normalizing by $n$ and applying the limit $n\rightarrow\infty$, we arrive at the desired outer bound.

\subsection{Proof of Theorem \ref{Kdetthm}}  \label{p1}
\begin{figure}
\center
\includegraphics[width = 4 in]{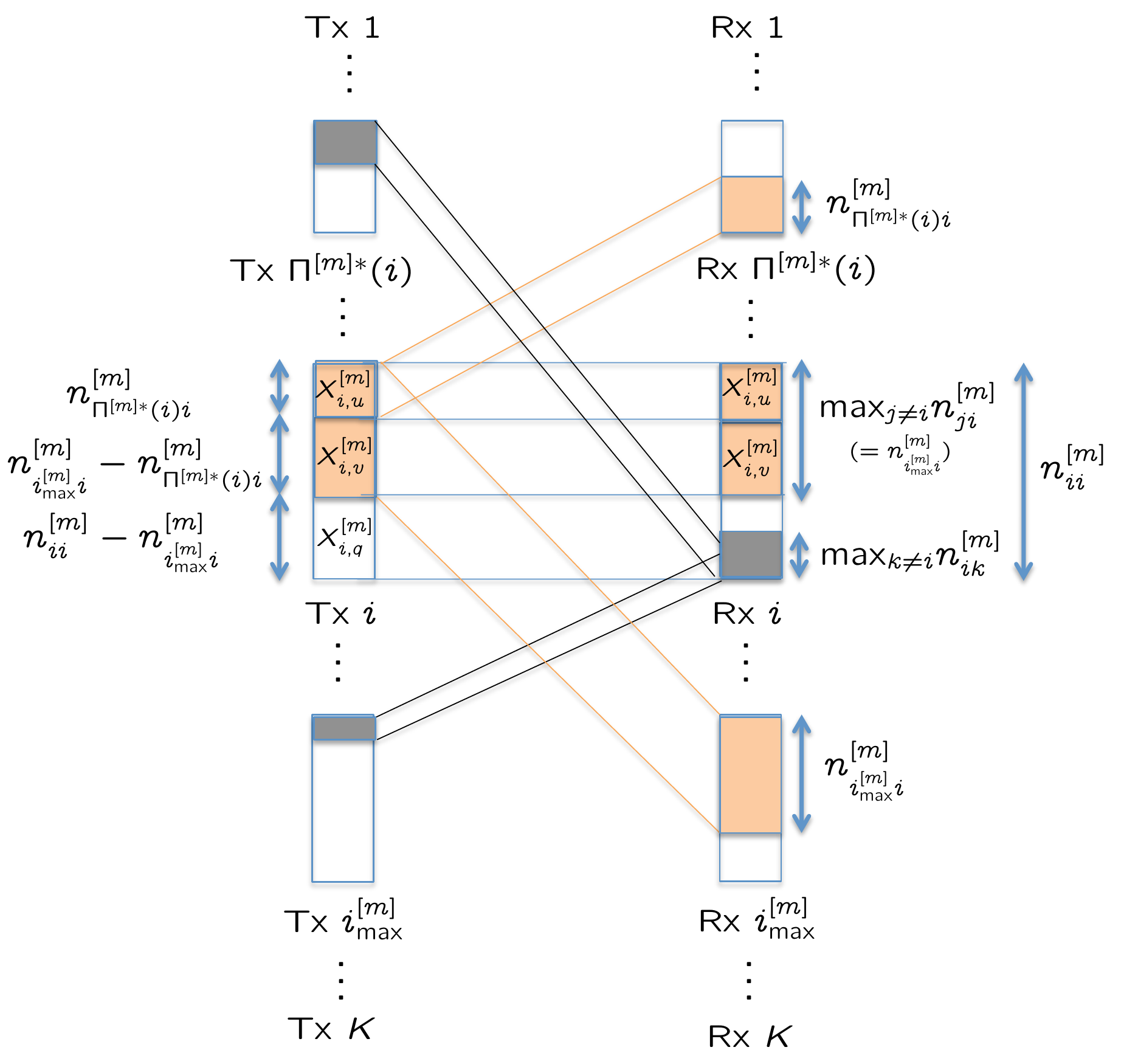}
\caption{\small The signal levels of Transmitter $i$ and Receiver $i$. As $n_{ii}^{[m]} \geq \max_{j\neq i} n_{ji}^{[m]} + \max_{k \neq i} n_{ik}^{[m]}$, the signal levels that cause interference ($X_{i,u}^{[m]}, X_{i,v}^{[m]}$) suffer no interference at the desired receiver.}
\label{weak}
\end{figure}

Corollary \ref{col:decompose} provides the achievable rate $\sum_{m=1}^M \left[ \sum_{i=1}^K n_{ii}^{[m]} - w(\Pi^{[m]*})\right]$ by separate TIN over each sub-channel. We only need to prove that it is an outer bound, under the assumption that each sub-channel is invertible. Consider the optimal cyclic partition for each sub-channel. Then by definition, $w(\Pi^{[m]*}) = \sum_{i=1}^K n_{\Pi^{[m]*} (i) i}$. We define $i_{\max}^{[m]} \triangleq {\mbox{argmax}}_ {j \neq i} ~n_{ji}^{[m]}$ to be the user that receives the most interference from Transmitter $i$ in sub-channel $m$. Writing the binary expansion of the channel input, 
\begin{eqnarray}
X_i^{[m]} 
&=& \sum_{b = 1}^{n_{ii}^{[m]}} X_{i,(b)}^{[m]} 2^{-b} \\
&=& \underbrace{ \sum_{b = 1}^{n_{\Pi^{[m]*} (i)i}^{[m]}} X_{i,(b)}^{[m]} 2^{-b} }_{\triangleq X_{i,u}^{[m]}} + \underbrace{ \sum_{b = n_{\Pi^{[m]*} (i)i}^{[m]}+1}^{n_{i_{\max}^{[m]}i}^{[m]}} X_{i,(b)}^{[m]} 2^{-b} }_{\triangleq X_{i,v}^{[m]}}   + \underbrace{ \sum_{b = n_{i_{\max}^{[m]}i}^{[m]}+1}^{n_{ii}^{[m]}} X_{i,(b)}^{[m]} 2^{-b} }_{\triangleq X_{i,q}^{[m]}}\\
&=&  X_{i,u}^{[m]} +  X_{i,v}^{[m]} +  X_{i,q}^{[m]}
\end{eqnarray}
where $X_{i,u}^{[m]}, X_{i,v}^{[m]},  X_{i,q}^{[m]}$ are the bits that interfere at Receiver $\Pi^{[m]*} (i)$, the other bits that interfere at Receiver $i^{[m]}_{\max}$ and the remaining input bits, respectively (see Figure \ref{weak}). We use ${\bf X}_{i,u}$ to denote the stack of $X_{i,u}^{[m]}$ for all sub-channels, i.e., ${\bf X}_{i,u} = [X_{i,u}^{[1]}, \ldots, X_{i,u}^{[M]}]$. Similar notation is used for  ${\bf X}_{i,v}$ with $v$ replacing $u$. 

Give ${\bf X}_{i,u}, {\bf X}_{i,v}$ as side information from a genie to Receiver $i$. Then from Fano's inequality, we have
\begin{eqnarray}
&& n(R_i - \epsilon) \notag \\
&\leq& I(W_i; {\bf Y}_i^n,  {\bf X}_{i,u}^n, {\bf X}_{i,v}^n) \\
&=& H({\bf Y}_i^n,  {\bf X}_{i,u}^n, {\bf X}_{i,v}^n) - H( {\bf Y}_i^n,  {\bf X}_{i,u}^n, {\bf X}_{i,v}^n | W_i) \\
&=& H({\bf X}_{i,u}^n | {\bf X}_{i,v}^n) + H({\bf X}_{i,v}^n) + H({\bf Y}_i^n | {\bf X}_{i,u}^n, {\bf X}_{i,v}^n) - H( {\bf Y}_i^n| W_i) \\
&\leq& H({\bf X}_{i,u}^n | {\bf X}_{i,v}^n) + n\sum_{m=1}^{M}( n_{i_{\max}^{[m]}i}^{[m]} -  n_{\Pi^{[m]*} (i)i}^{[m]}) + n\sum_{m=1}^{M}( n_{ii}^{[m]} - n_{i_{\max}^{[m]}i}^{[m]} ) - H( {\bf Y}_i^n | W_i) \label{detn} \\
&=& H({\bf X}_{i,u}^n | {\bf X}_{i,v}^n) - H( {\bf Y}_i^n | W_i) + n\sum_{m=1}^{M}( n_{ii}^{[m]} - n_{\Pi^{[m]*} (i)i}^{[m]} ) \label{detri}
\end{eqnarray}
where the second term in (\ref{detn}) follows from the fact that the entropy of $X_{i,v}^{[m]}$ is smaller than the number of bits therein  and the third term in (\ref{detn}) is due to the property that the signal levels in ${\bf Y}_i$ that receive ${\bf X}_{i,u}, {\bf X}_{i,v}$ do not suffer interference (see Figure \ref{weak}), because of each sub-channel is TIN optimal.

Adding (\ref{detri}) for $i \in \{1,\ldots,K\}$, we have
\begin{eqnarray}
\sum_{i=1}^K n( R_i - \epsilon) &\leq& \sum_{i=1}^K H({\bf X}_{i,u}^n | {\bf X}_{i,v}^n) - \sum_{i=1}^K H( {\bf Y}_i^n | W_i) + n\sum_{i=1}^K \sum_{m=1}^{M}( n_{ii}^{[m]} - n_{\Pi^{[m]*} (i)i}^{[m]} )  \\
&\leq& H({\bf X}_{1,u}^n, \ldots, {\bf X}_{K,u}^n| {\bf X}_{1,v}^n, \ldots, {\bf X}_{K,v}^n) - \sum_{i=1}^K H( {\bf Y}_i^n | W_i, {\bf X}_{1,v}^n, \ldots , {\bf X}_{K,v}^n) \notag \\
&& ~+ n \sum_{m=1}^{M}( \sum_{i=1}^K n_{ii}^{[m]} -  \sum_{i=1}^K n_{\Pi^{[m]*} (i)i}^{[m]} ) \label{ind} \\
&=& H({\bf X}_{1,u}^n, \ldots, {\bf X}_{K,u}^n| {\bf X}_{1,v}^n, \ldots, {\bf X}_{K,v}^n) - H( {\bf Y}_{1,u}^n, \ldots, {\bf Y}_{K,u}^n | {\bf X}_{1,v}^n, \ldots , {\bf X}_{K,v}^n) \notag \\
&& +~ n \sum_{m=1}^{M}\left[ \sum_{i=1}^K n_{ii}^{[m]} -  w(\Pi^{[m]*}) \right] \label{yu} \\
&=&n \sum_{m=1}^{M}\left[ \sum_{i=1}^K n_{ii}^{[m]} -  w(\Pi^{[m]*}) \right] \label{fin}
\end{eqnarray}
where (\ref{ind}) follows from the independence of ${\bf X}_i$ and the fact that conditioning does not increase entropy. The second term of (\ref{yu}) follows from the definition of $Y_{k,u}^{[m]} = \sum_{i=1,i \neq k}^K 2^{ n_{ki}^{[m]} } X_{i,u}^{[m]}$ and the fact that given the desired message $W_k$ and $X_{i,v}^{[m]}, X_{i,p}^{[m]}$, the only thing left in $Y_{k}^{[m]}$ is $Y_{k,u}^{[m]}$. (\ref{fin}) is due to the invertibility assumption. Normalizing (\ref{fin}) by $n$ and applying the limit $n\rightarrow\infty$, we arrive at the desired outer bound. 

\subsection{Proof of Theorem \ref{thm}} \label{p3}
As separate TIN achieves GDoF $\sum_{m=1}^M \left[ \sum_{i=1}^K \alpha_{ii}^{[m]} - w(\Pi^{[m]*}) \right]$, we prove that this is an outer bound. The proof is similar to that for the ADT deterministic model with the difference that the input of the Gaussian network has average power constraint 1. 
For sub-channel $m$, consider the optimal cyclic partition $\Pi^{[m]*}$ with weight $w(\Pi^{[m]*}) = \sum_{i=1}^K \alpha_{\Pi^{[m]*} (i)i}^{[m]}$.
Let us define $i_{\max}^{[m]}$ to be the user that receives the strongest interference from Transmitter $i$ over sub-channel $m$,  i.e., $i_{\max}^{[m]} \triangleq{\mbox{argmax}}_ {j \neq i} ~\alpha_{ji}^{[m]}$. 
Writing the binary expansion of the channel input,
\begin{eqnarray}
X_i^{[m]} 
&=& \mbox{sign}(X_i^{[m]}) \sum_{b = -\infty}^{\infty} X_{i,(b)}^{[m]} 2^{-b} \\
&=& \underbrace{ \mbox{sign}(X_i^{[m]}) \sum_{b = -\infty}^{0} X_{i,(b)}^{[m]} 2^{-b} }_{\triangleq X_{i,p}^{[m]}} +  \underbrace{ \mbox{sign}(X_i^{[m]}) \sum_{b = 1}^{n_{\Pi^{[m]*} (i)i}^{[m]} } X_{i,(b)}^{[m]} 2^{-b} }_{\triangleq X_{i,u}^{[m]}} \notag \\
&& + \underbrace{ \mbox{sign}(X_i^{[m]}) \sum_{b = n_{\Pi^{[m]*} (i)i}^{[m]} +1}^{n_{i^{[m]}_{\max}i}^{[m]}} X_{i,(b)}^{[m]} 2^{-b} }_{\triangleq X_{i,v}^{[m]}}   + \underbrace{\mbox{sign}(X_i^{[m]}) \sum_{b = n_{i^{[m]}_{\max}i}^{[m]} +1}^{\infty} X_{i,(b)}^{[m]} 2^{-b} }_{\triangleq X_{i,q}^{[m]}} \label{iq} \\
&=&  X_{i,p}^{[m]} + X_{i,u}^{[m]} +  X_{i,v}^{[m]} +  X_{i,q}^{[m]}
\end{eqnarray}
where $X_{i,p}^{[m]}, X_{i,u}^{[m]}, X_{i,v}^{[m]},  X_{i,q}^{[m]}$ are the bits that have power more than 1,  the bits that interfere at Receiver $\Pi^{[m]*} (i)$, the other interfering bits that appear at Receiver $i^{[m]}_{\max}$ and the remaining input bits that may only appear at the desired receiver, respectively. 
${\bf X}_{i,u}$ is used to denote $[X_{i,u}^{[1]}, \ldots, X_{i,u}^{[M]}]$. Similar notations are used for  ${\bf X}_{i,p}, {\bf X}_{i,v}$ with $p,v$ replacing $u$, respectively.

We borrow a lemma from \cite{Bresler_Tse} to bound the entropy of ${\bf X}_{i,p}$, the bits that have peak power more than 1. Intuitively, it means that those bits only have bounded entropy, thus limited influence on capacity.
\begin{lemma}
(Lemma 6 in \cite{Bresler_Tse}) The following bound on the entropy holds:  $H({\bf X}_{i,p}^n) \leq 2nM$.
\end{lemma}
For a proof, we refer the readers to \cite{Bresler_Tse}.

Giving ${\bf X}_{i,u}, {\bf X}_{i,v}$ and ${\bf X}_p \triangleq ({\bf X}_{1,p}, \ldots ,{\bf X}_{K,p})$ as side information from a genie to Receiver $i$, we have
\begin{eqnarray}
&& n(R_i - \epsilon) \notag\\
&\leq& I(W_i; {\bf {Y}}_i^n, {\bf X}_{i,u}^n, {\bf X}_{i,v}^n, {\bf X}_{p}^n) \\
&=& I(W_i; {\bf X}_{p}^n) + I(W_i;  {\bf X}_{i,u}^n, {\bf X}_{i,v}^n | {\bf X}_{p}^n) + I(W_i; {\bf {Y}}_i^n | {\bf X}_{i,u}^n, {\bf X}_{i,v}^n,{\bf X}_{p}^n) \\
&=&  \underbrace{ H( {\bf X}_{p}^n)}_{\leq nO(1)} - \underbrace{ H( {\bf X}_{p}^n | W_i) }_{\geq 0} + H({\bf X}_{i,u}^n, {\bf X}_{i,v}^n | {\bf X}_{p}^n)  - \underbrace{ H({\bf X}_{i,u}^n, {\bf X}_{i,v}^n | {\bf X}_{p}^n,W_i) }_{=0} \notag \\
&& ~+ h({\bf {Y}}_i^n |  {\bf X}_{i,u}^n, {\bf X}_{i,v}^n,{\bf X}_{p}^n) - \underbrace{ h( {\bf {Y}}_i^n | {\bf X}_{i,u}^n, {\bf X}_{i,v}^n, {\bf X}_{p}^n,W_i) }_{= h( {\bf {Y}}_i^n| W_i)}  \label{le} \\
&\leq& H({\bf X}_{i,u}^n | {\bf X}_{i,v}^n, {\bf X}_{p}^n) + H({\bf X}_{i,v}^n | {\bf X}_{p}^n) + h({\bf {Y}}_i^n | {\bf X}_{i,u}^n, {\bf X}_{i,v}^n, {\bf X}_{p}^n) - h( {\bf {Y}}_i^n| W_i) + nO(1) \\
&\leq& H({\bf X}_{i,u}^n | {\bf X}_{i,v}^n, {\bf X}_{p}^n) + n\sum_{m=1}^{M}(  n_{i^{[m]}_{\max}i}^{[m]} - n_{\Pi^{[m]*} (i)i}^{[m]})  \notag \\ && ~+ n\sum_{m=1}^{M} \frac{1}{2} \log \Big[ 2\pi e \big( \sum_{j\neq i} P^{\alpha_{ij}^{[m]}} + P^{\alpha_{ii}^{[m]}}2^{ -2 n_{i^{[m]} _{\max}i}^{[m] } }\big) \Big]
- h( {\bf {Y}}_i^n | W_i) + nO(1)  \label{gdetn}  \\
&\leq& H({\bf X}_{i,u}^n | {\bf X}_{i,v}^n, {\bf X}_{p}^{n})  + n\sum_{m=1}^{M}\left[ \frac{1}{2}( \alpha_{i_{\max}^{[m]} i}^{[m]} - \alpha_{\Pi^{[m]*} (i)i}^{[m]} )\log P + 1 \right]  \notag \\
&& ~+ n\sum_{m=1}^{M} \frac{1}{2} \log \Big[ 2\pi e \big(K P^{\alpha_{ii}^{[m]} - \alpha_{i^{[m]} _{\max}i}^{[m] } }\big) \Big] - h( {\bf {Y}}_i^n | W_i) + nO(1)  \label{po} \\
&=& H({\bf X}_{i,u}^n | {\bf X}_{i,v}^n, {\bf X}_{p}^{n}) - h( {\bf {Y}}_i^n | W_i) + n\sum_{m=1}^{M}\frac{1}{2}( \alpha_{ii}^{[m]} - \alpha_{\Pi^{[m]*} (i)i}^{[m]} )\log P  + nO(1) \label{last}
\end{eqnarray}
where we use Lemma 1 in the first term of (\ref{le}). The third term in (\ref{gdetn}) is due to the fact that the differential entropy of a random variable is maximized by Gaussian distribution given the covariance constraint, and conditioning on $X_{i,p}^{[m]}, X_{i,u}^{[m]}, X_{i,v}^{[m]}$, the magnitude of desired input is smaller than $2^{-n^{[m]}_{i^{[m]}_{\max} i}}$ (see (\ref{iq})). All the remaining interfering input has power constraint 1. In (\ref{po}), we use $n_{ki}^{[m]} = \lfloor \frac{1}{2} \alpha_{ki}^{[m]} \log_2 P\rfloor \subset (\frac{1}{2} \alpha_{ki}^{[m]} \log_2 P - 1, \frac{1}{2} \alpha_{ki}^{[m]} \log_2 P ]$ and the TIN optimality condition such that $\alpha_{ij}^{[m]} \leq \alpha_{ii}^{[m]} - \alpha_{i^{[m]} _{\max}i}^{[m] }$.

Adding (\ref{last}) for $i \in \{1,\ldots,K\}$, we have
\begin{eqnarray}
&& \sum_{i=1}^K n( R_i - \epsilon) \notag\\
&\leq& \sum_{i=1}^K \left[ H({\bf X}_{i,u}^n | {\bf X}_{i,v}^n, {\bf X}_{p}^{n}) -  h( {\bf {Y}}_i^n | W_i) \right] + n\sum_{i=1}^K \sum_{m=1}^{M}\frac{1}{2}( \alpha_{ii}^{[m]} - \alpha_{\Pi^{[m]*} (i)i}^{[m]} )\log P  + nO(1) \\
&\leq& H({\bf X}_{u}^n | {\bf X}_{v}^n, {\bf X}_{p}^{n}) - \sum_{i=1}^K h( {\bf {Y}}_i^n | W_i, {\bf X}_{v}^n, {\bf X}_{p}^{n}) \notag \\
&& +~ \frac{n}{2} \log P \sum_{m=1}^{M}  \left( \sum_{i=1}^K \alpha_{ii}^{[m]} - \sum_{i=1}^K \alpha_{\Pi^{[m]*} (i)i}^{[m]} \right) + nO(1)  \label{gfin} \\
&\leq& H({\bf X}_{u}^n | {\bf X}_{v}^n, {\bf X}_{p}^{n}) - h( {\bf {Y}}_{u}^n | {\bf X}_{v}^n, {\bf X}_{p}^{n}) + \frac{n}{2} \log P \sum_{m=1}^{M}  \left[ \sum_{i=1}^K \alpha_{ii}^{[m]} - w(\Pi^{[m]*}) \right]  + nO(1)  \label{y} \\
&\leq& \frac{n}{2} \log P  \sum_{m=1}^{M}  \left[ \sum_{i=1}^K \alpha_{ii}^{[m]} - w(\Pi^{[m]*}) \right]  + no(\log P) \label{inv}
\end{eqnarray}
where in (\ref{gfin}), ${\bf X}_u$ is the collection of ${\bf X}_{i,u}$ for all users, i.e., ${\bf X}_{u} = ({\bf X}_{1,u}, \ldots, {\bf X}_{K,u})$. Similar notations are used for ${\bf X}_{v}$ and ${\bf Y}_{u}$. The second term of (\ref{y}) is due to the definition that $Y_{i,u}^{[m]} = \sum_{j=1,j\neq i}^K h_{ij}^{[m]} X_{j,u}^{[m]} + Z_{i}^{[m]}$ and given the desired message $W_i$ and $X_{j,v}^{[m]}, X_{j,p}^{[m]}$, the only thing left in the received signal $Y_{i}^{[m]}$ is $Y_{i,u}^{[m]}$. (\ref{inv}) is due to the invertibility assumption and is derived as follows.
\begin{eqnarray}
&& H({\bf X}_{u}^n | {\bf X}_{v}^n, {\bf X}_{p}^{n}) - h( {\bf {Y}}_{u}^n | {\bf X}_{v}^n, {\bf X}_{p}^{n}) \notag \\
&=& H({\bf X}_{u}^n | {\bf {Y}}_{u}^n, {\bf X}_{v}^n, {\bf X}_{p}^{n}) - h( {\bf {Y}}_{u}^n | {\bf {X}}_{u}^n, {\bf X}_{v}^n, {\bf X}_{p}^{n}) \\
&\leq& \sum_{t=1}^n H({\bf X}_{u}(t) | {\bf {Y}}_{u}(t) , {\bf X}_{v}(t), {\bf X}_{p}(t)) + n o(\log P) \\
&\leq& \sum_{t=1}^n H({\bf X}_{u}(t) | {\bf {Y}}_{u}(t)) + n o(\log P) \\
&\leq& \sum_{t=1}^n \sum_{m=1}^M H({\bf X}_{u}^{[m]}(t) | {\bf {Y}}_{u}^{[m]}(t)) + n o(\log P) \\
&\leq& Mn o(\log P) + n o(\log P) = n o(\log P) 
\end{eqnarray}
where   the last inequality follows from the definition of invertibility as stated in (\ref{equ:ginv}).

Normalizing (\ref{inv}) by $\frac{1}{2} n \log P$ and letting first $n$ and then $P$ approach infinity, we obtain the matching outer bound and complete the proof.


\section{Discussions}
In the context of $K$ user parallel Gaussian interference networks when each sub-channel satisfies the TIN optimal condition of Geng et al., we show that separate TIN over each sub-channel is optimal under a mild condition from the perspective of sum-GDoF. The main message is that the simple ADT deterministic model is still very insightful for the optimality of TIN, because TIN is robust enough to not be sensitive to the details that are not captured by the ADT deterministic model. 

\bibliographystyle{IEEEtran}
\bibliography{Thesis}
\end{document}